\documentclass[%
 reprint,
superscriptaddress,
 amsmath,amssymb,
 aps,
]{revtex4-2}
\usepackage{xcolor}
\usepackage{graphicx}
\usepackage{dcolumn}
\usepackage{bm}
\usepackage{gensymb}
\usepackage{soul}
\usepackage{multirow}
\usepackage[T1]{fontenc}
\usepackage{booktabs}
\usepackage{xr}

\begin{document}

\preprint{APS/123-QED}

\title{Amplification of interlayer exciton emission in twisted
WSe$_2$/WSe$_2$/MoSe$_2$ heterotrilayers}

\author{Chirag C. Palekar}\email{c.palekar@tu-berlin.de}
\affiliation{Institute of Solid State Physics, Technische Universität Berlin, 10623 Berlin, Germany}

\author{Paulo E. {Faria~Junior}} \email{fariajunior.pe@gmail.com} 
\affiliation{Institute of Theoretical Physics, University of Regensburg, 93040 Regensburg, Germany}

\author{Barbara Rosa}\email{rosa@physik.tu-berlin.de} 
\affiliation{Institute of Solid State Physics, Technische Universität Berlin, 10623 Berlin, Germany}

\author{Frederico B. Sousa}
\affiliation{Departamento de Física, Universidade Federal de Minas Gerais, Belo Horizonte, Minas Gerais 30123-970}

\author{Leandro M. Malard}
\affiliation{Departamento de Física, Universidade Federal de Minas Gerais, Belo Horizonte, Minas Gerais 30123-970}

\author{Jaroslav Fabian}
\affiliation{Institute of Theoretical Physics, University of Regensburg, 93040 Regensburg, Germany}

\author{Stephan Reitzenstein}
\affiliation{Institute of Solid State Physics, Technische Universität Berlin, 10623 Berlin, Germany}

\setcounter{figure}{0}
\renewcommand{\figurename}{\textbf{Figure}}
\renewcommand{\thefigure}{\arabic{figure}}

\begin{abstract}

Transition metal dichalcogenide (TMDC) heterostructures have unique properties that depend on the twisting angle and stacking order of two or more monolayers. However, their practical applications are limited by the low photoluminescence yield of interlayer excitons. This limits the use of layered 2D materials as a versatile platform for developing innovative optoelectronic and spintronic devices. In this study, we report on the emission enhancement of interlayer excitons in multilayered-stacked monolayers through the fabrication of heterotrilayers consisting of WSe$_2$/WSe$_2$/MoSe$_2$ with differing twist angles. Our results show that an additional WSe$_2$ monolayer introduces new absorption pathways, leading to an improvement in the emission of interlayer excitons by more than an order of magnitude. The emission boost is affected by the twist angle, and we observe a tenfold increase in the heterotrilayer area when there is a $44\degree$ angle between the WSe$_2$ and MoSe$_2$ materials, as opposed to their heterobilayer counterparts. Furthermore, using density functional theory, we identify the emergence of new carrier transfer pathways in the three-layer sample which extends the current understanding of 2D semiconducting heterostructures. In addition, our research provides a viable way to significantly enhance the emission of interlayer excitons. The emission enhancement of interlayer excitons is significant not only for studying the fundamental properties of interlayer excitons, but also for enabling optoelectronic applications that utilize engineered 2D quantum materials with high luminescence yield.
\end{abstract}


\maketitle

\section{\label{sec:int}Introduction}

Interlayer excitons in heterostructures of transition metal dichalcogenides (TMDCs) have recently been identified as a newly discovered class of strongly bound quasiparticles \cite{Seyler2019Nature,Alexeev2019,Tran2019,Miao2021,Li2021,Wang2021}. These emerging materials are of great interest for both fundamental physics and practical applications, as they facilitate the investigation of intriguing optoelectronic properties and the development of innovative nanophotonic devices. A unique aspect of these TMDC heterostructures is that their electronic and optical properties can be controlled by conveniently changing the twist angle at which the layers are stacked. The twist angle determines the periodicity and confinement potential of the moiré superlattice, subsequently impacting the exciton physics of TMDC heterostructures \cite{Brem2020}. Therefore, remarkable many-body phenomena, including exciton interactions, band-gap renormalization, and Mott transition, have emerged in those systems \cite{Regan2020, Troue2023, Miao2021}. In addition, the moiré pattern associated with atomic reconstruction \cite{Rosenberger2020, VanWinkle2023, Zhao2023NatNano} and domain formation \cite{Sung2020,Parzefall2021} has a significant impact on the electronic band structure, which results in the formation of flat minibands \cite{Li2021,Alexeev2019} and hybridized excitonic states \cite{Alexeev2019}, both of which are often observed in heterostructures composed of TMDCs.

\begin{figure*}[!htb]
 \centering
 \includegraphics[width=0.85\textwidth]{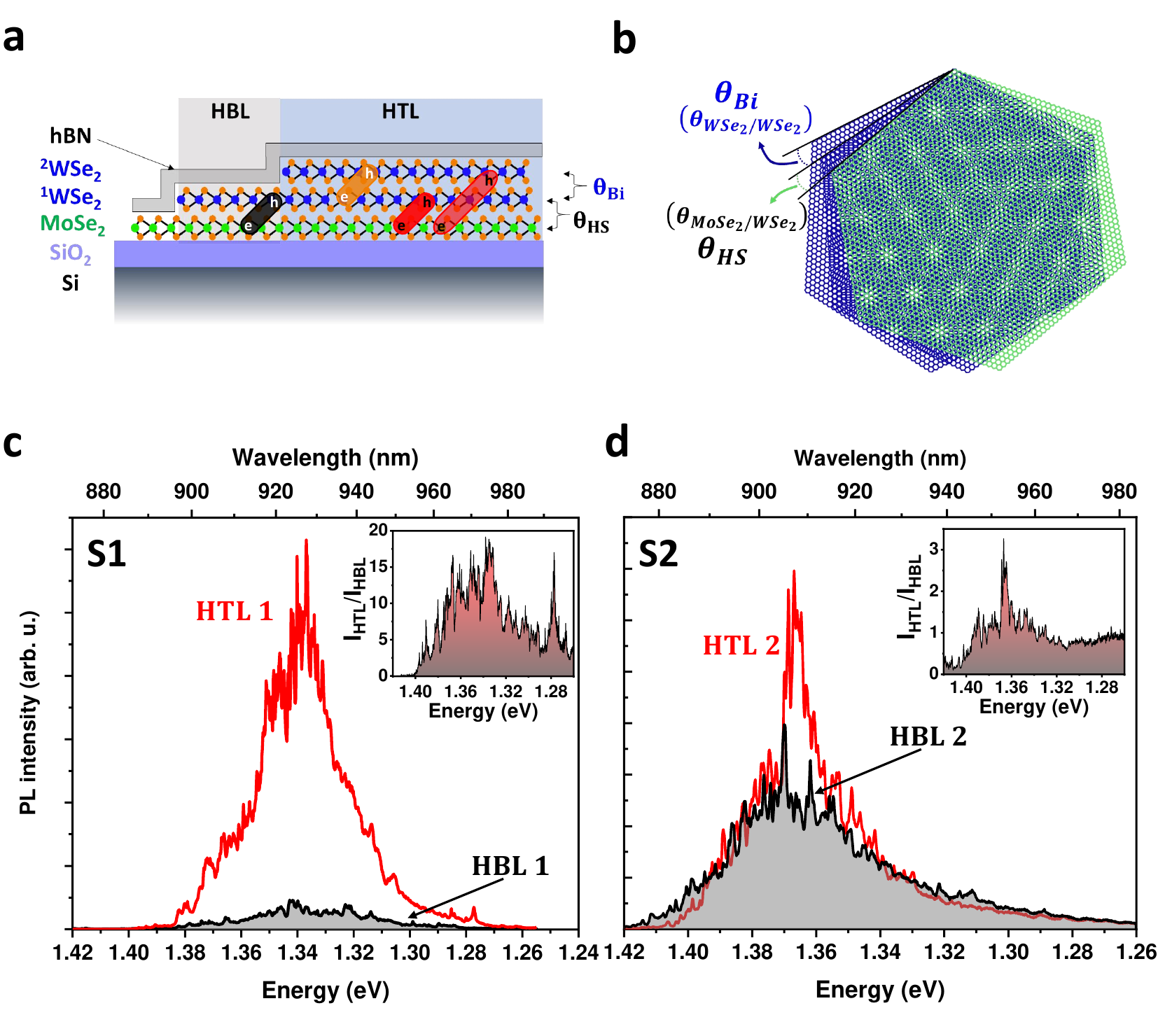}
 \caption{\textbf{Twisted $^2$WSe$_2$/$^1$WSe$_2$/MoSe$_2$ heterotrilayers.}({\textbf a}) Schematic representation of the HTL system showcasing the stacking order and the substrate configuration along with potential interlayer and momentum-indirect exciton formation between the layers.({\textbf b}) A simplified illustration of the HTL system (WSe$_2$ and MoSe$_2$ are shown in blue and green respectively) showcasing complex moiré superlattices with the twist angle ($\theta_{Bi}$ and $\theta_{HS}$) between the layers. ({\textbf {c,d}}) Photoluminescence spectra of the interlayer exciton transition highlighted in ({\textbf c}) from S1  and ({\textbf d}) from S2 measured at 4 K under resonant excitation at the WSe$_2$ intralayer exciton energy. The iX emission from the HTL (HBL) region is plotted in red (black) lines. The twist angle for the MoSe$_2$/WSe$_2$ heterostructure ($\theta_{HS}$) in S1 (S2) is $44\degree$ $\pm$ $3\degree$ ($58\degree$) and the twist angle for stacked WSe$_2$ MLs ($\theta_{Bi}$) is close to $0\degree$ ($57\degree$). The inset in {\textbf {c,d}} shows the ratio of measured intensity from HBL and HTL regions (I$_{HTL}$/I$_{HBL}$) of both samples demonstrating enhancement factor exceeding 15 and 3 for S1 and S2, respectively. }
 \label{fig1}
\end{figure*}

Among the various 2D semiconducting twisted bilayers presented in the literature, the WSe$_2$/MoSe$_2$ heterostructure exhibits a lattice mismatch of approximately $0.1\%$, resulting in a moiré length of dozens of nanometers \cite{Andersen2021, Wang2021}. It also features a type-II band alignment that accommodates interlayer excitons (iXs) with emission below $1.4$ eV at cryogenic temperatures, when electrons and holes bound by Coulomb forces occupying the conduction band (CB) and valence band (VB), respectively, in separate monolayers \cite{ Rivera2015, Nayak2017, Nagler2017NatComm, Gillen2018, Seyler2019Nature,Andersen2021, Zhao2023NatNano}. These characteristics can together form trapping sites that behave as twist-angle-controlled quantum-dot-like confinement potentials for excitons \cite{Brem2020}.

Nevertheless, despite the ultrafast transfer of interlayer charges between the conduction and valence bands of the corresponding MoSe$_2$ and WSe$_2$ \cite{Zimmermann2020, Rivera2015}, both  intralayer photoluminescence (PL) intensity for the monolayers (ML) and the interlayer excitonic emission for the heterostructures exhibits a notable low PL yield, which proves to be a significant bottleneck for opto-electronic applications. The low luminescence output of these heterostructures can be attributed to various factors, such as the non-direct alignment of bands at the K-point of the Brillouin zone, the dependence of layer separation on the twist angle, as well as the spatial separation of electron and hole wave functions, which leads to lessened coupling strength \cite{Nayak2017}. Therefore, it is essential to investigate TMDC heterostructures that display exceptional optical features in combination with powerful PL emission.

In this work, we study artificially stacked heterobilayers (HBL) and heterotrilayers (HTL) based on WSe$_2$ and MoSe$_2$ twisted MLs. Prior research on HTL systems include either MLs stacked with other natural bilayers \cite{Chen2022, Forg2021NatComm} or HTL systems in a sandwich-like configuration (ML$_1$/ML$_2$/ML$_1$) with arbitrary twist angles \cite{Bai2022,Slobodkin2020,Lian2023}. Importantly, the influence of the twist angle on the optical characteristics of the HTL system remains unclear and poorly explored. In sharp contrast to previous research, we are investigating WSe$_2$/WSe$_2$/MoSe$_2$ HTL systems in which MLs are mechanically stacked and twisted.

In our advanced nanofabrication procedure, we stacked each layer separately, controlling the twist angle between all constituent MLs. The stacking arrangement for our hole transport HTL system consists of ML $^1$WSe$_2$/ML $^2$WSe$_2$/ML MoSe$_2$ covered by a thin (< 8 nm) layer of hBN, as depicted in Fig.~\ref{fig1}a. We developed HTL systems and analyzed the effect of the twist angle on band gap renormalization in individual monolayers and the optoelectronic response of TMDC heterostructures in detail. We examined two samples, S1 and S2, with twist angles between the individual WSe$_2$ MLs ($\theta_{Bi}$) of 0$\degree$ and 57$\degree$, respectively. The 2H stacked configuration for the twisted heterobilayer closely resembles the twist angle found in the WSe$_2$/MoSe$_2$ heterostructure ($\theta_{HS}$), which is $44\degree$ $\pm$ $3\degree$ and 58$\degree$, respectively. The twist angles were determined by  polarization resolved second harmonic generation (SHG) measurements \cite{Malard2013}. Additional details can be found in the methods section C and supplementary information (SI) section 1. The HBL and HTL regions are labeled as HBL1 (HBL2) and HTL1 (HTL2) for samples S1 (S2), respectively. Figure~\ref{fig1}({\textbf a}) displays the stacked arrangement of our artificially twisted heterostructure made of $^2$WSe$_2$/$^1$WSe$_2$/MoSe$_2$. Figure \ref{fig1}({\textbf b}) shows the HTL and HBL areas are made up of $^2$WSe$_2$/$^1$WSe$_2$/MoSe$_2$ and $^1$WSe$_2$/MoSe$_2$, respectively, along with ($\theta_{Bi}$) and ($\theta_{HS}$). \\

Upon comparing the HBL and HTL regions of samples S1 and S2 using micro-photoluminescence ($\mu$PL), reflectance contrast, photoluminescence excitation (PLE), and degree of circular polarization (DCP) measurements, we observed a significant difference in iXs emission at cryogenic temperatures. Notably, the HTL system showed a considerable enhancement in emission, up to 10 times, compared to its HBL region for S1. Further, we use density functional theory (DFT) to model and assess the electronic band structures of our twisted HBL and HTL systems. The calculated dipole matrix elements between valence and conduction band states at the K-point give insight into potential transitions in twisted heterostructure systems. Our findings showcase diverse opportunities that arise from manipulating the properties of TMDC heterostructures through twist-angle engineering. In this study, we present a new and efficient method for manipulating the number of monolayers to regulate the interlayer excitonic response and PL yield in twisted heterostructures. Our investigation of heterotrilayers extends beyond simple heterobilayers and creates new avenues for exploring the parameter space of TMDC quantum materials. This includes analyzing the stacking order and twist angles between all constituent layers, thereby offering exciting prospects for further research and potential applications.

\section{Results}

\subsection{Amplification of interlayer exciton photoluminescence} 

To investigate iXs in different HBL and HTL environments of samples S1 and S2, we utilized PL spectroscopy at 4 K with resonant excitation at the WSe$_2$ intralayer transition around 1.72 eV (additional details in Methods section B). Figures~\ref{fig1} (c, d) display the characteristic PL emission from iX in $^1$WSe$_2$/MoSe$_2$ HBLs, ranging from 1.3 to 1.4 eV \cite{Rivera2015, Seyler2019Nature, Tran2019, Nayak2017, Nagler2017NatComm, Zhao2023NatNano}. The greater physical separation between interlayers due to the large twist angle leads to significantly lower PL emission intensity of iX from the HBL1 (S1) region, as opposed to HBL2 (S2). The separation between layers, which is dependent on the twist angle, impedes swift charge transfer. As a result, the population of iX in HBL1 decreases. Our DFT calculations (see Methods sections D and E) establish that the oscillator strength of interlayer excitons diminishes as the twist angle increases, thereby causing a reduction in emission.

The red curve in the HTL1 region of S1 shows a significant increase in PL intensity when compared to the black curve in the HBL1 region, as seen in Fig.~\ref{fig1}c. In Fig.~\ref{fig1}d, we see the PL response of S2, which has some improvement in iX PL, but not as noticeable as in S1. The comparison of iX emission in Fig.~\ref{fig1}c shows S1 to have a tenfold increase in integrated intensities. The ratio of HBL/HTL iX's emission intensity shows a more than 15-fold increase for S1 and roughly 3-fold increase for S2 at distinct emission energies. We also gathered numerous PL spectra from various points in S1 and S2, revealing a significant increase in PL output that remained consistent for the HTL area. This can be seen in Section 3 of the SI. After studying the light emissions from both samples (see Fig.~\ref{fig1}c and Fig.~\ref{fig1}d) and considering how the different layers are arranged, we suggest that the extra layer, purposely twisted at angles around $0\degree$ or $60\degree$, creates more pathways for light emission during the transfer of energy between MoSe$_2$$\leftrightarrow$WSe$_2$ layers. On the other hand, emission intensity of HTL1 is still less than that of HTL2, specifically due to large momentum mismatch corresponding to large twist angle between $^1$WSe$_2$/MoSe$_2$ interface.  

\subsection{Influence of the twist angle on the interlayer exciton emission in heterotrilayers}

\begin{figure*}[t] 
 \centering
 \includegraphics[scale=0.6]{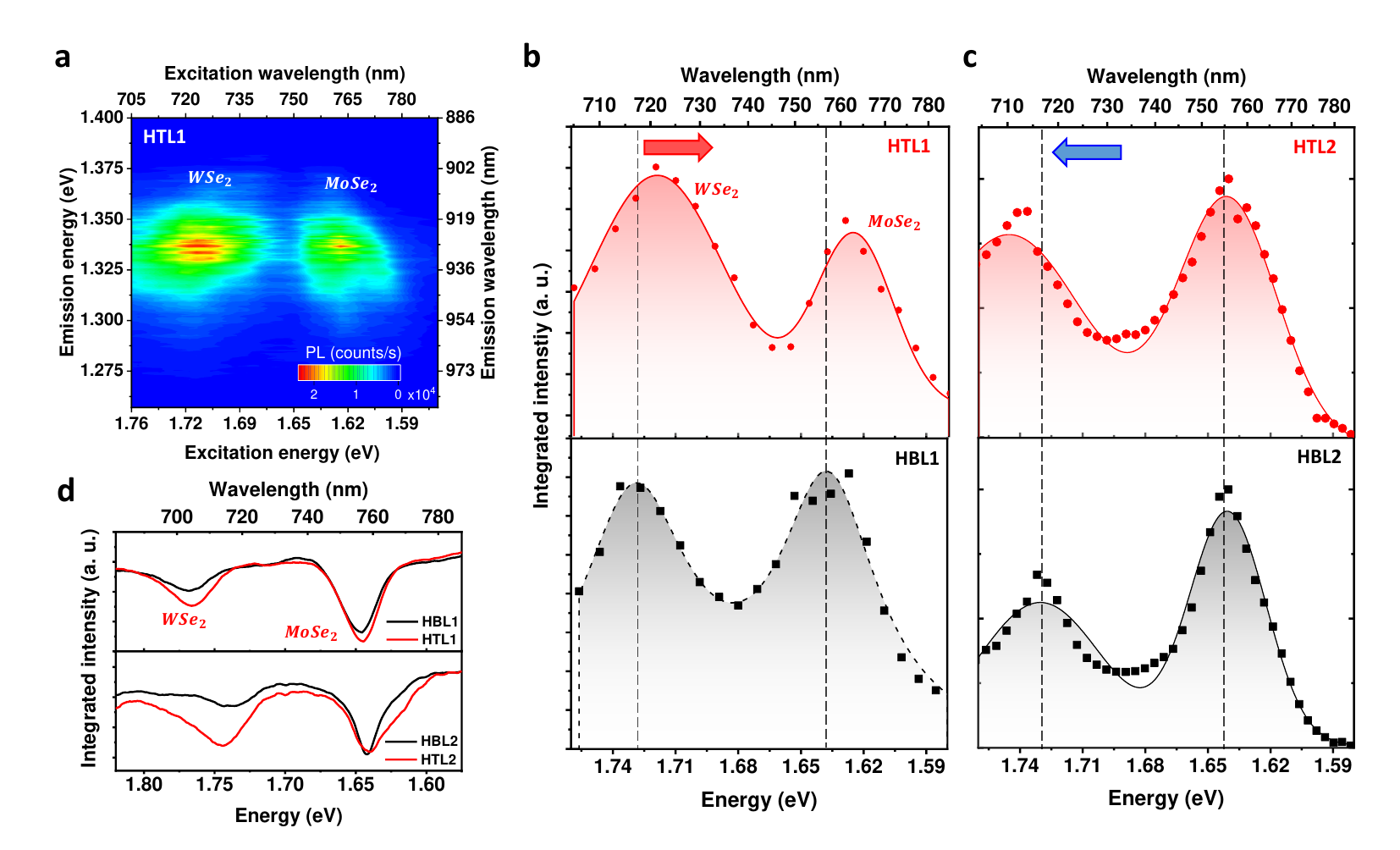}
 \caption{\textbf{Photoluminescence excitation spectroscopy of twisted heterobilayers and heterotrilayers.}({\textbf {a}}) False color PLE map of iX emission from the HTL1 region manifesting the intralayer exciton  resonances of WSe$_2$ and MoSe$_2$. ({\textbf {b, c}}) Direct comparison between the PLE resonances recorded from all four studied regions of the two samples (S1, S2). The integrated intensity of the interlayer exciton as function of excitation energy is plotted in ({\textbf {b, c}}), upper panel, for HTL (red dots) and lower panel for HBL (black dots). Further the PLE response is fitted with a double Gaussian function (solid line with shaded area), see Table S1 for fit parameters. The vertical dashed black line helps to identify energy shifts (red and blue arrow) of PLE resonances. ({\textbf {d}}) Reflection contrast measurements on the S1 and S2 with all considered regions (black curve for HBL, red curve for HTL). In the HTL regions, the light absorption is stronger than in HBL regions particularly from A exciton of WSe$_{2}$ due to the additional WSe$_2$ ML. The energy shift of WSe$_{2}$ absorption in HTL1 and HTL2 region is attributed to twist angle induced hybridisation of intralayer exciton states \cite{Merkl2020}. }
 \label{fig2}
\end{figure*}

We performed PLE measurements to study the relationship between the twist angle of stacked WSe$_2$ homobilayers and the spin-valley physics of the HTLs. We created a false color PLE map that shows the iX intensity based on excitation wavelength (energy) at the HTL1 region (Fig.~\ref{fig2}a). We examined how WSe$_2$ and MoSe$_2$ contribute to forming the iX using PLE spectroscopy. By using a pulsed-laser to excite the samples from $1.75$~eV ($705$~nm) to $1.55$~eV ($790$~nm) at constant power, we were able to detect two separate PLE resonances which corresponded to the exciton A of both WSe$_2$ and MoSe$_2$ (refer to Fig.~\ref {fig2}a). Please see the Methods Section B for additional information.

For comparison, we gathered the PLE map of the HBL and HTL regions from both S1 and S2. The PLE maps for all regions are shown in the SI (see Fig. S3). We extracted the integrated intensity for the HTL (HBL) region as a function of energy from the measured PLE maps, which is presented in Figure~\ref {fig2}(b, c) for both samples. Additionally, we used a Gaussian function to fit the integrated intensity, with fitting parameters given in the SI (Table S1). The PLE scan reveals two resonances that match the intralayer exciton (WSe$_2$ and MoSe$_2$) of the constituent MLs, displaying a higher response in integrated intensity. This indicates not only an efficient absorption from each ML individually and improved gain from heterostructure but also efficient interlayer electron-hole tunneling between WSe$_2$/MoSe$_2$ HBL interface, which substantiates the interlayer PL emission in Fig.~\ref{fig1}(b, c).

Based on our analysis (see SI Table S1), we found that in the HBL1 region, the combined integrated intensities of WSe$_2$ and MoSe$_2$ are approximately equal, indicating similar contributions from both materials. In contrast, in the HTL1 region, we observe a significant difference between the intensities for WSe$_2$ and MoSe$_2$.  The PLE intensity of iX from the HTL1 area shows that the WSe$_2$ resonance is enhanced roughly by 2 times compared to the HBL1 area. Meanwhile, the PLE resonance intensity of MoSe$_2$ does not significantly change in both HTL and HBL for S1 (please see Table S1). A similar change in the integrated intensity aspect ratio of WSe$_2$ is observed for the HBL2 and HTL2 areas (please see Fig.~\ref{fig2}c and Table S1). Furthermore, the weight of each material's absolute PLE may differ from sample to sample, as indicated in the PLE literature \cite{Li2020, Hanbicki2018}. Our PLE findings suggest that the extra WSe$_2$ ML boosts light absorption and iX exciton population, leading to the enhanced iX emission demonstrated in the PL spectra in Fig.~\ref{fig1}(c,d). 

\begin{figure*}[t]
 \centering
 \includegraphics[scale=0.55]{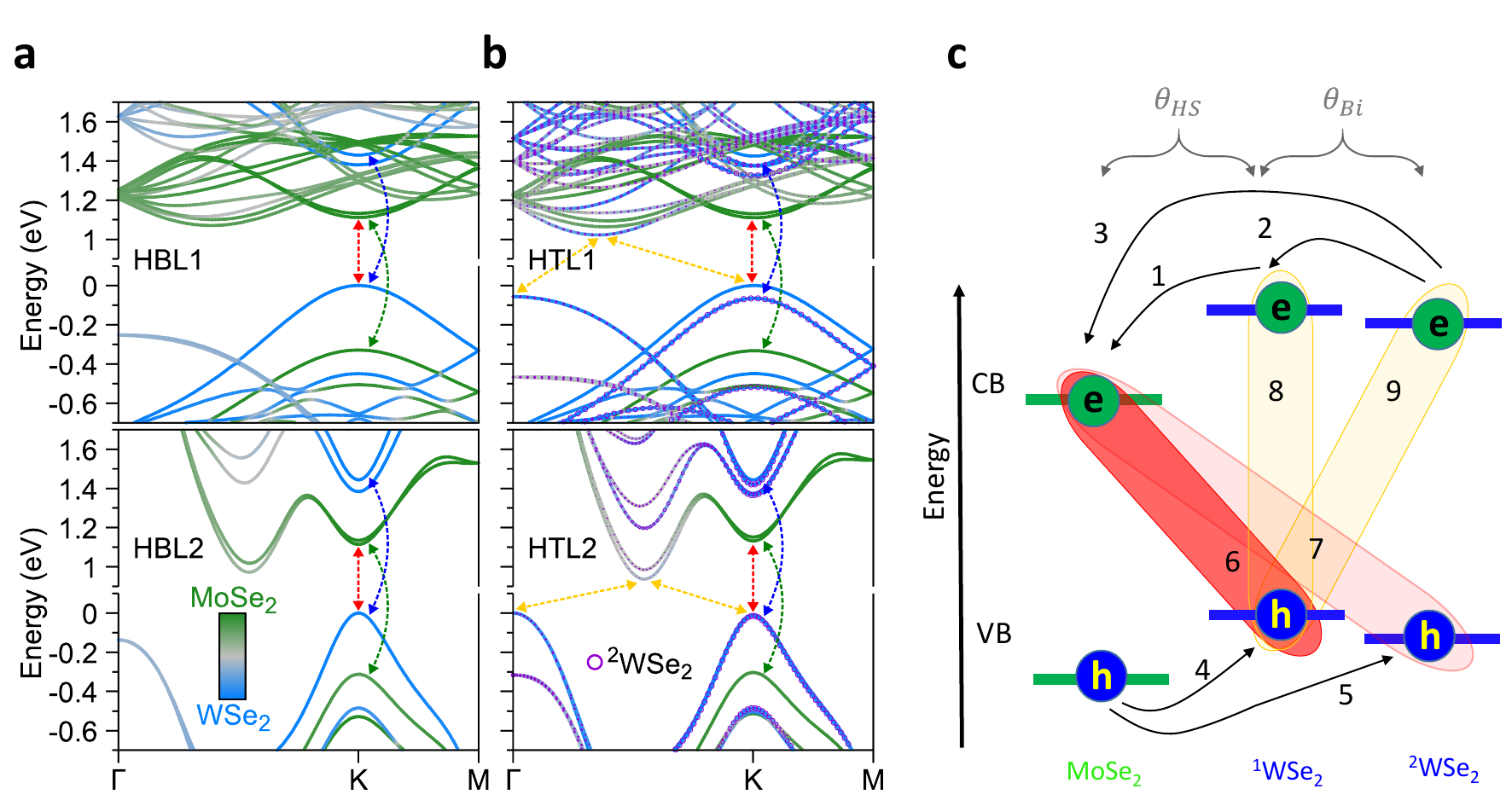}
 \caption{\textbf{Electronic structure of heterobilayers and heterotrilayers.} Band structure calculations of selected HBL ({\textbf {a}}) and HTL ({\textbf {b}}) systems, with color code representing MoSe$_2$ and WSe$_2$ layers. In panel ({\textbf {b}}), the top-most WSe$_2$ layer is identified with the circles (the size is proportional to the contribution atomic composition). Particularly, HBL1 refers to $\sim \! 38.2 \degree$ between MoSe$_2$ and $^1$WSe$_2$, HBL2 refers to the $60 \degree$ case with H$^{\textrm{h}}_{\textrm{h}}$ stacking, HTL1 refers to a R$^{\textrm{X}}_{\textrm{h}}$ stacking between $^2$WSe$_2$ and $^1$WSe$_2$, and HTL2 refers to a H$^{\textrm{h}}_{\textrm{h}}$ stacking between $^2$WSe$_2$ and $^1$WSe$_2$. The other investigated band structures are presented in the SI (see section IV). The arrows highlight the relevant optical transitions. 
 For intra- and inter-layer direct transitions at the K valley, the calculated momentum matrix elements are given in Tables~S3-S4 in the SI. ({\textbf {c}})The $^1$WSe$_2$ and $^2$WSe$_2$ conduction band minima are shifted indicating slight mismatch due to twist angle and effect of hybridization. The scattering paths (1-3 for electrons and 4-5 for holes) and optically active interlayer exciton transitions (6,7) along with momentum-indirect excitons in WSe$_2$ (8,9) are specified. }
 \label{fig3}
\end{figure*}

\begin{figure*}[t]
 \centering
 \includegraphics[scale=0.9]{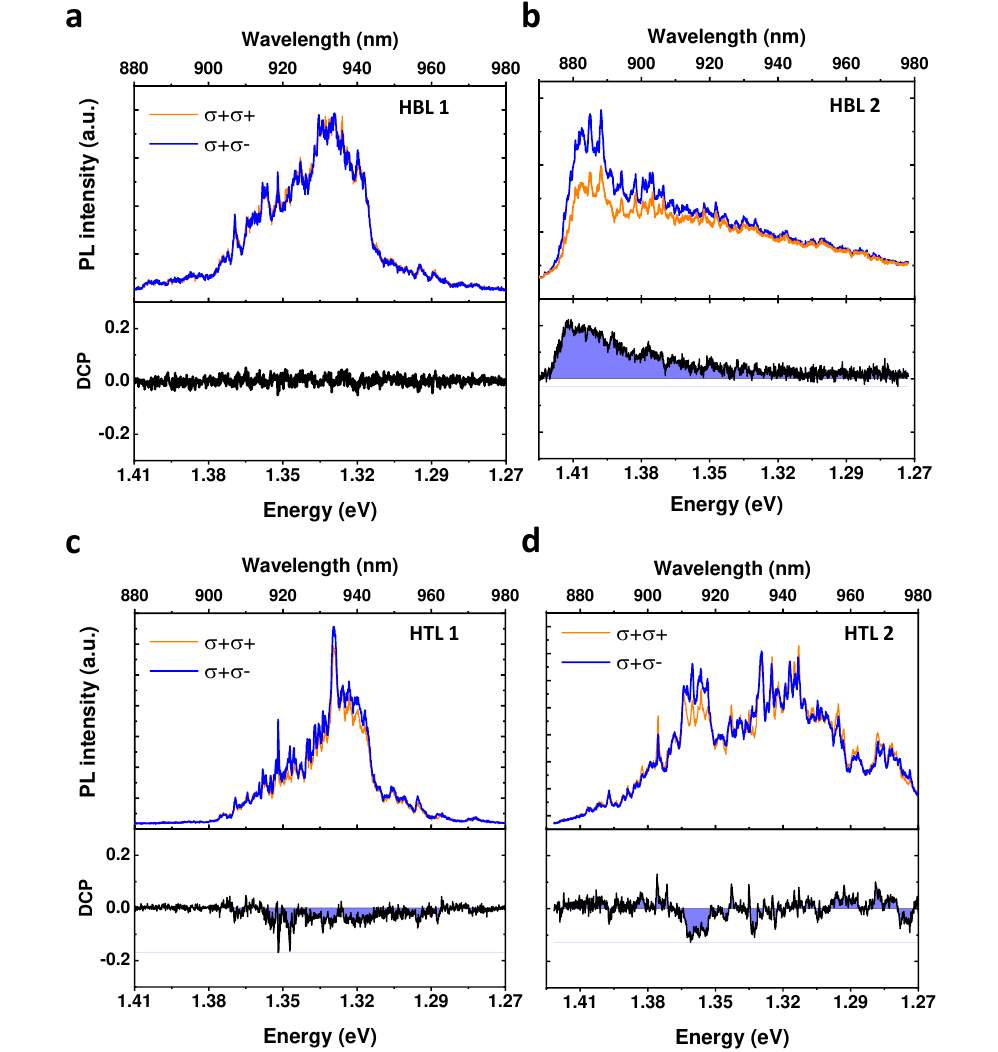}
 \caption{\textbf{Influence of the twist angle on polarization of interlayer exciton from twisted heterotrilayers.}({\textbf {a,b}}) Valley polarization of the iX from HBL region of S1 ({\textbf a}) and S2 ({\textbf b}). In the upper panel, we show measured right ($\sigma$$^+$) and left ($\sigma$$^-$) circularly polarized light is plotted which was obtained by using a pulsed laser with $\sigma$$^+$ polarization. The bottom panel shows the calculated DCP. ({\textbf {c,d}}) Valley polarization of the iX from HTL regions along with DCP in the bottom panel.}
 \label{fig4}
\end{figure*}

Interestingly, a clear redshift was recorded for the WSe$_{2}$ PLE resonance in the HTL1 region in comparison to the HBL1 region (see Fig.~\ref{fig2}b), whereas in S2, WSe$_{2}$ PLE resonance shows significant blueshift under the same comparison (Fig.~\ref{fig2}c). PLE results underline our assumption that the WSe$_2$ twisted angle of the top layer plays a crucial role in the band structure of the HTL samples. The strongly enhanced iX PL yield corresponds to the R-type stacking of the WSe$_2$ homobilayer, whereas the H-type stacking configuration only moderately alters the iX emission response. Since PLE addresses the absorption signatures of optically active material, we determine the actual absorption response through reflectance measurements on S1 and S2, presented in Fig.~\ref{fig2}d. The absorption features can be attributed to WSe$_2$ ($\sim 1.73$~eV) and MoSe$_2$ ($\sim 1.66$~eV) excitons. In addition, both samples show an enhanced absorption caused by the presence of the WSe$_2$ stacked bilayer, whereas MoSe$_2$ appears similar for HTL and HBL in S1 and S2. As shown in PLE results, a blueshift in absorption is also depicted in S2, as shown in Fig.~\ref{fig2}d. However, no relevant redshift was observed in sample S1. 

To explain our findings, we closely scrutinize the hybridization phenomenon in twisted bilayers, responsible for altering the band structure of stacked TMDC system \cite{Merkl2020, Lin2021}. Previously, Merkl \textit{et al.}\cite{Merkl2020} and Lin \textit{et al.}\cite{Lin2021} investigated the angle dependence of intra- and interlayer excitons in WSe$_2$ homobilayers, where an intra- and interlayer hybridization effect was found to be responsible for altering the transitions in momentum space. As a result, the twist angle-dependent energy shifts, such as a blueshift for WSe$_2$ with 2H stacking and redshift with R-type stacking, are observed. The PLE and reflection contrast measurements are in good agreement with theoretical and experimental findings from Refs.\cite{Merkl2020, Lin2021}.

\section{Discussion} 

It is known that the twist angle strongly influences the band structure and, consequently, the exciton properties in TMDC heterostructures~\cite{Nayak2017, Lin2021, Merkl2020,  Zhang2020, Choi2021}. Therefore, due to the tunable nature of such heterostructures, the wave function overlap and hybridization of electronic states provide valuable insights into the intra- and inter-layer exciton properties. In order to provide fundamental microscopic insight into the PL and PLE enhancement, we performed first principles DFT calculations of relevant high-symmetry stacking configurations (see Fig.~S5 of the SI). Figs.~\ref{fig3}(a,b) show the calculated electronic band structures of the representative HBL and HTL systems, with the color code representing the layer localization of the wavefunction and the dashed arrows indicating the relevant optical transitions. For direct transitions at the K valleys, the calculated values of the dipole matrix elements are presented in Tables~S3, S4 of the SI. 

Our results indicate that the dipole matrix elements for intralayer exciton transitions are rather robust and nearly independent of the twist angle and stacking. Comparing the iXs for the HBL cases, the dipole matrix element of the iX in the 38.2$\degree$ system is one order of magnitude weaker, which is consistent with the reduced PL intensities observed experimentally at large twist angles\cite{liu2014evolution,zheng2015coupling,Nayak2017,shi2019twisted,Volmer2023npj,palekar2023twist}. For the iX species related to MoSe$_2-^2$WSe$_2$ transitions, our DFT calculations reveal that the dipole matrix elements are strongly dependent on the stacking configuration. For the energetically favorable HTL stackings presented in Fig.~\ref{fig3}b, the MoSe$_2-^2$WSe$_2$ iX has a comparable dipole matrix element to the MoSe$_2-^1$WSe$_2$ iX for the HTL2 case, but it shows a decrease of 2 orders of magnitude for the HTL1 case. To complement our analysis for direct iXs, we evaluated their binding energies, presented in Table~S6 of the SI. We found that the binding energy of the MoSe$_2-^2$WSe$_2$ iX is $\sim$8 meV smaller than the MoSe$_2-^1$WSe$_2$ iX (which is nearly independent of the stacking details since the effective masses are essentially the same, as shown in Table S5). Combining the energy separation of the WSe$_2$ valence bands (Table S2 of the SI) and the iX binding energies, we predict the MoSe$_2-^2$WSe$_2$ iX energy to be $\sim$15 (74) meV above the MoSe$_2-^1$WSe$_2$ iX in the HTL2 (HTL1) case.

We now turn to the momentum-indirect transitions\cite{kunstmann2018momentum} indicated by the orange lines in the HTL calculations of Fig.~\ref{fig3}b. In HBL MoSe$_2$/WSe$_2$ samples, transitions involving the Q (the conduction band minima between $\Gamma$ and K) and $\Gamma$ points seem to be strongly suppressed and not visible in the vast majority of the studied samples, with strong support from magneto-optical characterizations highlighting the nature of direct K-K interlayer excitons via their g-factors (see Refs.\cite{Wozniak2020PRB, FariaJunior2023} and references therein). In HTL systems, however, Q and $\Gamma$ points may become relevant since the energy bands at these points acquire an increased contribution of the WSe$_2$ character. This effect can be seen in Fig.~\ref{fig3}b by the distinct blue color appearing at the $\Gamma$ point of HTL1 and HTL2 (accompanied by an increase in the energy) and in the lowest conduction band in the Q point of HTL1. For clarity, Figs.~S8,S9 in the SI present the contribution of the different layers for the lowest conduction band and top valence bands in an alternative way to further support our analysis. Thus, momentum-indirect transitions with intra- or inter-layer characters originating from the WSe$_2$ bilayer region can also influence the observed PL enhancement, particularly for the HTL1 case in which the direct MoSe$_2-^1$WSe$_2$ iX is further away in energy and has a weaker dipole matrix element.

Based on our experiments and DFT calculations, we first explain the iX emission measured at the HBL region of S1 considering the optical transitions presented in the SI, Table~S3. The PLE response shown in Fig.~\ref{fig2}b supports the PL enhancements based on the additional absorbed light, as this results in an increased bright iX population in the HTL1 system due to rapid charge transfer between the layers. However, the DFT calculations suggest that there is very weak direct interaction between $^2$WSe$_2$ and MoSe$_2$ ML (see vW2$\leftrightarrow$cMo transition in Table S3 for HTL1). Therefore, the increased exciton population, hybridization, and additional exciton scattering paths due to $^2$WSe$_2$ in HTL systems are mainly responsible for the iX emission enhancement. For better understanding, possible scattering paths and transitions are shown using simple schematics in Fig.~\ref{fig3}c between the WSe$_2$ and MoSe$_2$ ML in the HTL1 region. Besides, exciton emission with z-polarization (dipole oriented out of heterostructure plane) was observed for TMDC ML, referred to as dark-gray exciton \cite{Jiang2018, Wang2017}. Moreover, it was shown experimentally in the HBL system that the dark-gray exciton, with the aid of scattering and spin flipping, can become an optically active iX \cite{Jiang2018}. Such a conversion can occur following dark-to-bright-to-iX exciton scattering paths, partially contributing to the emission enhancement. 

Considering S2, the dipole matrix elements for iX transitions (path 6), between adjacent $^1$WSe$_2$ and MoSe$_2$ monolayers (vW1$\leftrightarrow$cMo), associated with in-plane ($x,y$) polarization are relatively similar for HBL and HTL regions. According to our DFT calculations, presented in Fig.~\ref{fig3} and Table S3 for the HTL2 system, a direct transition between the $^2$WSe$_2$ ML and the bottom MoSe$_2$ ML (vW2$\leftrightarrow$cMo) are shown to be possible (path 7 in Fig.~\ref{fig3}). However, the values of dipole matrix elements for such transitions vW2$\leftrightarrow$ cMo are relatively smaller than that of vW1$\leftrightarrow$cMo because of $^1$WSe$_2$ being in between $^2$WSe$_2$ and MoSe$_2$ as this increases the interlayer separation.  
Moreover, our calculations indicate that the interlayer excitons originating from the MoSe$_2$ layer to the $^2$WSe$_2$ can lie close in energy but have strongly suppressed optical transitions. This result is in line with recent findings in trilayer systems composed of a natural bilayer MoSe$_2$ stacked on monolayer WSe$_2$ \cite{Forg2021NatComm}.

To investigate the influence of intervalley and dark exciton scattering on iX, as the role of moiré/atomic reconstruction, as shown in Ref.\cite{Zhao2023NatNano} we measured the polarization resolved PL from HBL and HTL regions (see Fig.~\ref{fig4}) with excitation resonant to the A exciton of WSe$_2$ being circularly polarized ($\sigma_+$). Following the excitation, only left ($\sigma_-$) and right ($\sigma_+$) circularly emission was selected and recorded. The degree of circular polarization, DCP $= (\sigma_- - \sigma_+)/(\sigma_- + \sigma_+) $, is shown in the lower panel of Fig.~\ref{fig4}. Large twist angles in heterobilayers result in a longer lifetime of iX due to intervalley interactions \cite{Choi2021}, dark exciton scattering \cite{Jiang2018}, and phonon-assisted momentum-indirect transitions (such as K-Q, K-$\Gamma$) \cite{Forg2021NatComm}, as shown in Fig.~\ref{fig3}b. Specifically, spin mixing, intervalley, and dark exciton scattering result in DCP losses, which is seen in HBL1 compared to HBL2. A similar response has been reported in Ref.\cite{Seyler2019Nature}. On the other hand, for small twist angles, the iX transition is nearly momentum-direct at K-K points, which avoids the inter-valley scattering and preserves the DCP (see Fig.~\ref{fig3}b). The positive value of DCP in S2 is consistent with the interlayer triplet state from H$^{\textrm{h}}_{\textrm{h}}$ stacking \cite{Zhao2023NatNano}. The DCP losses support our initial attribution in which valley spin mixing and dark exciton scattering play a crucial role in iX emission from HBL systems with large twist angles. Besides, introducing $^2$WSe$_2$ ML increases not only the bright excitons but also the dark exciton population, which further boosts iX formation in HTL systems through charge transfer and additional allowed transitions shown in Fig.~\ref{fig3}b. The marginal revival of the DCP observed in HTL1 region (Fig.~\ref{fig4}c) can be attributed to relatively larger energy splitting between $^2$WSe$_2$ and $^1$WSe$_2$ at VB(K) compared to HTL2, as shown in Fig.~\ref{fig3}. However, the loss of DCP in HTL2 systems (see Fig.~\ref{fig4}d) further solidifies our reasoning of iX enhancement based on valley spin mixing and additional scattering paths. Lastly, based on our experimental and theoretical findings, we highlight the preferential stacking alignment of $R_{h}^{M}$ for $^2$WSe$_2$/$^1$WSe$_2$ in the S1 system, which provides substantial iX emission enhancement even in the HBL system with large twist angles.

\section{Conclusion}

In conclusion, our combined theoretical and experimental study shows that mechanically stacked WSe$_2$/WSe$_2$/MoSe$_2$ HTL systems with varying twist angles can lead to iX PL enhancement by up to 10 times in the HTL regions, compared to their HBL counterparts. Previous work on WSe$_2$/MoSe$_2$ HBL system with large twist angles showcased very low iX intensity. This makes such HBL systems impracticable for detailed investigation and applications due to their large momentum mismatch as well as large interlayer separation, causing the reduced iX population. In contrast, our findings introduce an efficient and powerful platform for enhancing the iX population and emission intensity. The recorded 10 times enhancement of iX emission results from increased exciton population due to intentionally added WSe$_2$ on top of twisted WSe$_2$/MoSe$_2$ HBL system. In a nutshell, our investigation demonstrates that the iX PL intensity in the twisted WSe$_2$/MoSe$_2$ systems can be enhanced, and controlled, by simply adding the additional layer of WSe$_2$ with R- or H- type configuration. Further, detailed DFT calculation shed more light on the significant interaction between three layers contributing to the iX formation and relaxation, enhancing the iX emission from HTL regions. This fundamental study of excitons in the HTL system deepens the current understanding of the physics of twisted TMDC heterostructures and paves the way for further intricate experiments and theoretical work.

\section{Methods}

\subsection{Sample fabrication}

The multilayered TMDC
samples were fabricated using mechanical exfoliation\cite{Novoselov2004} and the dry-transfer method\cite{Castellanos-Gomez2014}. For exfoliation of the MLs, commercially available crystals of MoSe$_2$ and WSe$_2$ are used along with Netto blue tape and Polydimethylsiloxane (PDMS). Bulk crystals of MoSe$_2$ and WSe$_2$ are mechanically exfoliated on the PDMS gel strip with the help of scotch tape. Next, the individual MLs are identified by optical contrast microscopy and PL spectroscopy at room temperature. Single crystalline MLs of MoSe$_2$ and WSe$_2$ are selected by observing the formed straight edges, at 60$\degree$ or 120$\degree$, as an indication of crystal axes. For the first sample (S1) fabrication, the MoSe$_2$ ML was transferred onto a $180$~nm layer of SiO$_2$ deposited on Si substrate. During the stacking process, the WSe$_2$ ML was aligned to the edge of MoSe$_2$ ML, addressing 0$\degree$ or 60$\degree$ twist angles. The WSe$_2$ ML was partially brought in contact with the MoSe$_2$ monolayer until touching it, thus intentionally being retrieved from the substrate in order to split the WSe$_2$ ML into two parts. To fabricate the trilayer, the remaining WSe$_2$ ML, attached to the PDMS, is transferred onto the WSe$_2$/MoSe$_2$ heterobilayer. The nominal twist angle between the two WSe$_2$ was therefore 0$\degree$. To achieve top encapsulation, an exfoliated thin ($\sim 5$~nm) hBN layer is transferred onto the heterotrilayer for both samples. Following a similar sample preparation method, a second sample (S2) was achieved; however, in this case, the second WSe$_2$ monolayer transferred was obtained using a new monolayer, and the angle between the WSe$_2$/WSe$_2$ bilayer was 60$\degree$. 

\subsection{Photoluminescence spectroscopy}

The PL spectroscopy was used to measure the excitonic emission from the MLs and heterostructure systems at room temperature for identification of MLs as well as at low temperature to study the iX emission from HTL and HBL regions. For ML identification, continues wave excitation laser was used with wavelength at 532 nm. Low temperature measurements were conducted at 4 K using close-cycle cryostat equipped with a high numerical aperture (NA = 0.81) objective lens. For low temperature measurements, a wavelength tunable picosecond mode lock pulsed laser was used as source of excitation to perform the photoluminescence excitation measurements. For the PLE measurements, the iX emission intensity was recorded by tuning the laser from 705 nm to 790 nm while keeping the excitation power constant over the tuning range. 
 
\subsection{Second harmonic generation}
The twist angle between the constituent layers of the samples S1 and S2 was determined by performing polarization resolved second harmonic generation (SHG) measurements \cite{xubaka2013}. The intensity as a function of the excitation laser polarization is recorded from the ML transferred on the substrate, which was excited with a linearly polarized picosecond mode-locked laser.  The TMDC ML is excited with linearly polarised femtosecond mode-locked laser with wavelength of 1313 nm. Then, the SHG intensity (at 556) as function of excitation laser polarisation is recorded from the ML transferred on the substrate.The characteristic intensity maxima show a six-fold symmetry, and each maximum indicates the armchair direction on the hexagonal crystal lattice of the associated TMDC ML. Comparing the SHG response from the constituent ML of the heterostructure system, the twist angle between the ML can be determined with high accuracy. To differentiate between R- or H-type stacking, we measured the SHG response of the HBL and the HTL regions. Reduction in intensity of SHG signal from the HBL region with respect to MLs indicates H-type stacking. Similarly, for stacked WSe2$_2$ MLs, R-type stacking results in higher SHG signal than H-type stacking as it leads to restored inversion symmetry \cite{Hsu2014SecondHG}. The overall measurement and fitting error is in the range of 1$\degree$ to 3$\degree$.

\subsection{Density Functional Theory}

We evaluate the electronic and optical properties of the HBL WSe$_2$/MoSe$_2$ and HTL WSe$_2$/WSe$_2$/MoSe$_2$ based on DFT using the all-electron full-potential linearized augmented plane-wave (LAPW) method within the Wien2k code\cite{wien2k}. We employ the Perdew–Burke–Ernzerhof\cite{Perdew1996PRL} exchange-correlation functional with van der Waals interactions included via the D3 correction\cite{Grimme2010JCP}. The wave function expansion into atomic spheres takes into account orbital quantum numbers up to 10 and the plane-wave cut-off multiplied with the smallest atomic radii is set to 8. Spin–orbit coupling was included fully relativistically for core electrons, while valence electrons were treated within a second-variational procedure\cite{Singh2006} with the scalar-relativistic wave functions calculated in an energy window up to 2 Ry. Self-consistency was achieved using a two-dimensional Monkhorst–Pack k-grid with 15$\times$15 points and the convergence criteria of $10^{-6} \; e$ for the charge and $10^{-6} \; \textrm{Ry}$ for the energy were used. We considered an average lattice parameter of 3.2855 $\textrm{\AA}$ for MoSe$_2$ and WSe$_2$ which amounts to an average and negligible strain of $\sim \! 0.1 \%$\cite{FariaJunior2023}. A vacuum region of 20 $\textrm{\AA}$ was considered to avoid interaction between the heterostructures replicas. The atomic positions were relaxed in the out-of-plane direction using a force convergence threshold of $10^{-4}$ Ry/Bohr. For sample S1, we constructed a twisted supercell with an angle of $\sim \! 38.2 \degree$ between MoSe$_2$ and the first WSe$_2$ and investigated the effect of different R-type stackings between the first WSe$_2$ and second WSe$_2$ layer, namely the R$^{\textrm{M}}_{\textrm{h}}$ and R$^{\textrm{X}}_{\textrm{h}}$ stackings (R$^{\textrm{h}}_{\textrm{h}}$ is known to be energetically unfavorable\cite{FariaJunior2023}). For sample S2, we assumed the H$^{\textrm{h}}_{\textrm{h}}$ stacking between MoSe$_2$ and the first WSe$_2$ and investigated the effect of different H-type stackings (H$^{\textrm{h}}_{\textrm{h}}$, H$^{\textrm{M}}_{\textrm{h}}$, and H$^{\textrm{X}}_{\textrm{h}}$) between the first WSe$_2$ and second WSe$_2$ layer. The details of the different stackings, as well as the structural parameters, are given in the SI. The dipole matrix element between valence and conduction band states at the K point are calculated as $\frac{\hbar}{m_0} \left\langle v, K \left| \vec{p} \cdot \hat{\alpha} \right| c, K \right\rangle$ for the LAPW basis set\cite{Draxl2006CPC}, in which $\hat{\alpha}$ econdes the light polarization ($\sigma^+$, $\sigma^-$, and $z$). See Table S2 in the SI.

\subsection{Exciton binding energies}

The calculated binding energies for the interlayer excitons were obtained using the effective Bethe-Salpeter equation\cite{Rohlfing1998PRL, Rohlfing2000PRB} formalism, assuming non-interacting parabolic bands for electrons and holes\cite{Ovesen2019CommPhys, FariaJunior2023TDM}. We consider the effective masses taken from DFT and the interlayer electrostatic potential obtained numerically by solving the Poisson equation in $k$-space, assuming the different regions to have dielectric constants of $\varepsilon(\textrm{MoSe}_2) = 16.8$ (\cite{Laturia2018npj2D}) and $\varepsilon(\textrm{WSe}_2) = 15.3$ (\cite{Laturia2018npj2D}), $\varepsilon(\textrm{SiO}_2) = 3.9$ (\cite{Berkelbach2013PRB}), and $\varepsilon(\textrm{hBN}) = 4.5$ (\cite{Stier2018PRL}), and the TMDCs to have an effective thickness of 6.72 (6.78) for MoSe$_2$ (WSe$_2$), taken as twice the value of the physical thickness of the TMDC\cite{Berkelbach2013PRB}. The calculated interlayer exciton potentials, effective masses, and binding energies are given in theSI.

\section*{Acknowledgements}
CCP, BR and SR acknowledge financial support of the Deutsche Forschungsgemeinschaft (DFG, German Research Foundation) via SPP 2244 (Project No. 443416027). PEFJ and JF acknowledge financial support of the DFG via SFB 1277 (Project-ID 314695032, projects B07 and B11), SPP 2244 (Project No. 443416183), and of the European Union Horizon 2020 Research and Innovation Program under Contract No. 881603 (Graphene Flagship). FPS and LMM acknowledge financial support from CNPq, CAPES, FINEP, FAPEMIG, INCT de Nanomateriais de Carbono a Rede Mineira de Materiais 2D.


\begin{thebibliography}{58}%
\makeatletter
\providecommand \@ifxundefined [1]{%
 \@ifx{#1\undefined}
}%
\providecommand \@ifnum [1]{%
 \ifnum #1\expandafter \@firstoftwo
 \else \expandafter \@secondoftwo
 \fi
}%
\providecommand \@ifx [1]{%
 \ifx #1\expandafter \@firstoftwo
 \else \expandafter \@secondoftwo
 \fi
}%
\providecommand \natexlab [1]{#1}%
\providecommand \enquote  [1]{``#1''}%
\providecommand \bibnamefont  [1]{#1}%
\providecommand \bibfnamefont [1]{#1}%
\providecommand \citenamefont [1]{#1}%
\providecommand \href@noop [0]{\@secondoftwo}%
\providecommand \href [0]{\begingroup \@sanitize@url \@href}%
\providecommand \@href[1]{\@@startlink{#1}\@@href}%
\providecommand \@@href[1]{\endgroup#1\@@endlink}%
\providecommand \@sanitize@url [0]{\catcode `\\12\catcode `\$12\catcode
  `\&12\catcode `\#12\catcode `\^12\catcode `\_12\catcode `\%12\relax}%
\providecommand \@@startlink[1]{}%
\providecommand \@@endlink[0]{}%
\providecommand \url  [0]{\begingroup\@sanitize@url \@url }%
\providecommand \@url [1]{\endgroup\@href {#1}{\urlprefix }}%
\providecommand \urlprefix  [0]{URL }%
\providecommand \Eprint [0]{\href }%
\providecommand \doibase [0]{https://doi.org/}%
\providecommand \selectlanguage [0]{\@gobble}%
\providecommand \bibinfo  [0]{\@secondoftwo}%
\providecommand \bibfield  [0]{\@secondoftwo}%
\providecommand \translation [1]{[#1]}%
\providecommand \BibitemOpen [0]{}%
\providecommand \bibitemStop [0]{}%
\providecommand \bibitemNoStop [0]{.\EOS\space}%
\providecommand \EOS [0]{\spacefactor3000\relax}%
\providecommand \BibitemShut  [1]{\csname bibitem#1\endcsname}%
\let\auto@bib@innerbib\@empty
\bibitem [{\citenamefont {Seyler}\ \emph {et~al.}(2019)\citenamefont {Seyler},
  \citenamefont {Rivera}, \citenamefont {Yu}, \citenamefont {Wilson},
  \citenamefont {Ray}, \citenamefont {Mandrus}, \citenamefont {Yan},
  \citenamefont {Yao},\ and\ \citenamefont {Xu}}]{Seyler2019Nature}%
  \BibitemOpen
  \bibfield  {author} {\bibinfo {author} {\bibfnamefont {K.~L.}\ \bibnamefont
  {Seyler}}, \bibinfo {author} {\bibfnamefont {P.}~\bibnamefont {Rivera}},
  \bibinfo {author} {\bibfnamefont {H.}~\bibnamefont {Yu}}, \bibinfo {author}
  {\bibfnamefont {N.~P.}\ \bibnamefont {Wilson}}, \bibinfo {author}
  {\bibfnamefont {E.~L.}\ \bibnamefont {Ray}}, \bibinfo {author} {\bibfnamefont
  {D.~G.}\ \bibnamefont {Mandrus}}, \bibinfo {author} {\bibfnamefont
  {J.}~\bibnamefont {Yan}}, \bibinfo {author} {\bibfnamefont {W.}~\bibnamefont
  {Yao}},\ and\ \bibinfo {author} {\bibfnamefont {X.}~\bibnamefont {Xu}},\
  }\bibfield  {title} {\bibinfo {title} {Signatures of moiré-trapped valley
  excitons in {MoSe}$_2$/{WSe}$_2$ heterobilayers},\ }\href
  {https://doi.org/10.1038/s41586-019-0957-1} {\bibfield  {journal} {\bibinfo
  {journal} {Nature}\ }\textbf {\bibinfo {volume} {567}},\ \bibinfo {pages}
  {66} (\bibinfo {year} {2019})}\BibitemShut {NoStop}%
\bibitem [{\citenamefont {Alexeev}\ \emph {et~al.}(2019)\citenamefont
  {Alexeev}, \citenamefont {Ruiz-Tijerina}, \citenamefont {Danovich},
  \citenamefont {Hamer}, \citenamefont {Terry}, \citenamefont {Nayak},
  \citenamefont {Ahn}, \citenamefont {Pak}, \citenamefont {Lee}, \citenamefont
  {Sohn}, \citenamefont {Molas}, \citenamefont {Koperski}, \citenamefont
  {Watanabe}, \citenamefont {Taniguchi}, \citenamefont {Novoselov},
  \citenamefont {Gorbachev}, \citenamefont {Shin}, \citenamefont {Fal'ko},\
  and\ \citenamefont {Tartakovskii}}]{Alexeev2019}%
  \BibitemOpen
  \bibfield  {author} {\bibinfo {author} {\bibfnamefont {E.~M.}\ \bibnamefont
  {Alexeev}}, \bibinfo {author} {\bibfnamefont {D.~A.}\ \bibnamefont
  {Ruiz-Tijerina}}, \bibinfo {author} {\bibfnamefont {M.}~\bibnamefont
  {Danovich}}, \bibinfo {author} {\bibfnamefont {M.~J.}\ \bibnamefont {Hamer}},
  \bibinfo {author} {\bibfnamefont {D.~J.}\ \bibnamefont {Terry}}, \bibinfo
  {author} {\bibfnamefont {P.~K.}\ \bibnamefont {Nayak}}, \bibinfo {author}
  {\bibfnamefont {S.}~\bibnamefont {Ahn}}, \bibinfo {author} {\bibfnamefont
  {S.}~\bibnamefont {Pak}}, \bibinfo {author} {\bibfnamefont {J.}~\bibnamefont
  {Lee}}, \bibinfo {author} {\bibfnamefont {J.~I.}\ \bibnamefont {Sohn}},
  \bibinfo {author} {\bibfnamefont {M.~R.}\ \bibnamefont {Molas}}, \bibinfo
  {author} {\bibfnamefont {M.}~\bibnamefont {Koperski}}, \bibinfo {author}
  {\bibfnamefont {K.}~\bibnamefont {Watanabe}}, \bibinfo {author}
  {\bibfnamefont {T.}~\bibnamefont {Taniguchi}}, \bibinfo {author}
  {\bibfnamefont {K.~S.}\ \bibnamefont {Novoselov}}, \bibinfo {author}
  {\bibfnamefont {R.~V.}\ \bibnamefont {Gorbachev}}, \bibinfo {author}
  {\bibfnamefont {H.~S.}\ \bibnamefont {Shin}}, \bibinfo {author}
  {\bibfnamefont {V.~I.}\ \bibnamefont {Fal'ko}},\ and\ \bibinfo {author}
  {\bibfnamefont {A.~I.}\ \bibnamefont {Tartakovskii}},\ }\bibfield  {title}
  {\bibinfo {title} {Resonantly hybridized excitons in moire superlattices in
  van der waals heterostructures},\ }\href
  {https://doi.org/10.1038/s41586-019-0986-9} {\bibfield  {journal} {\bibinfo
  {journal} {Nature}\ }\textbf {\bibinfo {volume} {567}},\ \bibinfo {pages}
  {81} (\bibinfo {year} {2019})}\BibitemShut {NoStop}%
\bibitem [{\citenamefont {Tran}\ \emph {et~al.}(2019)\citenamefont {Tran},
  \citenamefont {Moody}, \citenamefont {Wu}, \citenamefont {Lu}, \citenamefont
  {Choi}, \citenamefont {Kim}, \citenamefont {Rai}, \citenamefont {Sanchez},
  \citenamefont {Quan}, \citenamefont {Singh}, \citenamefont {Embley},
  \citenamefont {Zepeda}, \citenamefont {Campbell}, \citenamefont {Autry},
  \citenamefont {Taniguchi}, \citenamefont {Watanabe}, \citenamefont {Lu},
  \citenamefont {Banerjee}, \citenamefont {Silverman}, \citenamefont {Kim},
  \citenamefont {Tutuc}, \citenamefont {Yang}, \citenamefont {MacDonald},\ and\
  \citenamefont {Li}}]{Tran2019}%
  \BibitemOpen
  \bibfield  {author} {\bibinfo {author} {\bibfnamefont {K.}~\bibnamefont
  {Tran}}, \bibinfo {author} {\bibfnamefont {G.}~\bibnamefont {Moody}},
  \bibinfo {author} {\bibfnamefont {F.}~\bibnamefont {Wu}}, \bibinfo {author}
  {\bibfnamefont {X.}~\bibnamefont {Lu}}, \bibinfo {author} {\bibfnamefont
  {J.}~\bibnamefont {Choi}}, \bibinfo {author} {\bibfnamefont {K.}~\bibnamefont
  {Kim}}, \bibinfo {author} {\bibfnamefont {A.}~\bibnamefont {Rai}}, \bibinfo
  {author} {\bibfnamefont {D.~A.}\ \bibnamefont {Sanchez}}, \bibinfo {author}
  {\bibfnamefont {J.}~\bibnamefont {Quan}}, \bibinfo {author} {\bibfnamefont
  {A.}~\bibnamefont {Singh}}, \bibinfo {author} {\bibfnamefont
  {J.}~\bibnamefont {Embley}}, \bibinfo {author} {\bibfnamefont
  {A.}~\bibnamefont {Zepeda}}, \bibinfo {author} {\bibfnamefont
  {M.}~\bibnamefont {Campbell}}, \bibinfo {author} {\bibfnamefont
  {T.}~\bibnamefont {Autry}}, \bibinfo {author} {\bibfnamefont
  {T.}~\bibnamefont {Taniguchi}}, \bibinfo {author} {\bibfnamefont
  {K.}~\bibnamefont {Watanabe}}, \bibinfo {author} {\bibfnamefont
  {N.}~\bibnamefont {Lu}}, \bibinfo {author} {\bibfnamefont {S.~K.}\
  \bibnamefont {Banerjee}}, \bibinfo {author} {\bibfnamefont {K.~L.}\
  \bibnamefont {Silverman}}, \bibinfo {author} {\bibfnamefont {S.}~\bibnamefont
  {Kim}}, \bibinfo {author} {\bibfnamefont {E.}~\bibnamefont {Tutuc}}, \bibinfo
  {author} {\bibfnamefont {L.}~\bibnamefont {Yang}}, \bibinfo {author}
  {\bibfnamefont {A.~H.}\ \bibnamefont {MacDonald}},\ and\ \bibinfo {author}
  {\bibfnamefont {X.}~\bibnamefont {Li}},\ }\bibfield  {title} {\bibinfo
  {title} {Evidence for moiré excitons in van der waals heterostructures},\
  }\href {https://doi.org/10.1038/s41586-019-0975-z} {\bibfield  {journal}
  {\bibinfo  {journal} {Nature}\ }\textbf {\bibinfo {volume} {567}},\ \bibinfo
  {pages} {71} (\bibinfo {year} {2019})}\BibitemShut {NoStop}%
\bibitem [{\citenamefont {Miao}\ \emph {et~al.}(2021)\citenamefont {Miao},
  \citenamefont {Wang}, \citenamefont {Huang}, \citenamefont {Chen},
  \citenamefont {Lian}, \citenamefont {Wang}, \citenamefont {Blei},
  \citenamefont {Taniguchi}, \citenamefont {Watanabe}, \citenamefont {Tongay},
  \citenamefont {Wang}, \citenamefont {Xiao}, \citenamefont {Cui},\ and\
  \citenamefont {Shi}}]{Miao2021}%
  \BibitemOpen
  \bibfield  {author} {\bibinfo {author} {\bibfnamefont {S.}~\bibnamefont
  {Miao}}, \bibinfo {author} {\bibfnamefont {T.}~\bibnamefont {Wang}}, \bibinfo
  {author} {\bibfnamefont {X.}~\bibnamefont {Huang}}, \bibinfo {author}
  {\bibfnamefont {D.}~\bibnamefont {Chen}}, \bibinfo {author} {\bibfnamefont
  {Z.}~\bibnamefont {Lian}}, \bibinfo {author} {\bibfnamefont {C.}~\bibnamefont
  {Wang}}, \bibinfo {author} {\bibfnamefont {M.}~\bibnamefont {Blei}}, \bibinfo
  {author} {\bibfnamefont {T.}~\bibnamefont {Taniguchi}}, \bibinfo {author}
  {\bibfnamefont {K.}~\bibnamefont {Watanabe}}, \bibinfo {author}
  {\bibfnamefont {S.}~\bibnamefont {Tongay}}, \bibinfo {author} {\bibfnamefont
  {Z.}~\bibnamefont {Wang}}, \bibinfo {author} {\bibfnamefont {D.}~\bibnamefont
  {Xiao}}, \bibinfo {author} {\bibfnamefont {Y.-T.}\ \bibnamefont {Cui}},\ and\
  \bibinfo {author} {\bibfnamefont {S.-F.}\ \bibnamefont {Shi}},\ }\bibfield
  {title} {\bibinfo {title} {Strong interaction between interlayer excitons and
  correlated electrons in {WSe}2/{WS}2 moir{\'{e}} superlattice},\ }\href
  {https://doi.org/10.1038/s41467-021-23732-6} {\bibfield  {journal} {\bibinfo
  {journal} {Nature Communications}\ }\textbf {\bibinfo {volume} {12}},\
  \bibinfo {pages} {3608} (\bibinfo {year} {2021})}\BibitemShut {NoStop}%
\bibitem [{\citenamefont {Li}\ \emph {et~al.}(2021)\citenamefont {Li},
  \citenamefont {Hong}, \citenamefont {Gao}, \citenamefont {Lin}, \citenamefont
  {Lim}, \citenamefont {Lu}, \citenamefont {Wu}, \citenamefont {Liu},
  \citenamefont {Tateyama}, \citenamefont {Sakuma}, \citenamefont {Tsukagoshi},
  \citenamefont {Suenaga},\ and\ \citenamefont {Taniguchi}}]{Li2021}%
  \BibitemOpen
  \bibfield  {author} {\bibinfo {author} {\bibfnamefont {S.}~\bibnamefont
  {Li}}, \bibinfo {author} {\bibfnamefont {J.}~\bibnamefont {Hong}}, \bibinfo
  {author} {\bibfnamefont {B.}~\bibnamefont {Gao}}, \bibinfo {author}
  {\bibfnamefont {Y.-C.}\ \bibnamefont {Lin}}, \bibinfo {author} {\bibfnamefont
  {H.~E.}\ \bibnamefont {Lim}}, \bibinfo {author} {\bibfnamefont
  {X.}~\bibnamefont {Lu}}, \bibinfo {author} {\bibfnamefont {J.}~\bibnamefont
  {Wu}}, \bibinfo {author} {\bibfnamefont {S.}~\bibnamefont {Liu}}, \bibinfo
  {author} {\bibfnamefont {Y.}~\bibnamefont {Tateyama}}, \bibinfo {author}
  {\bibfnamefont {Y.}~\bibnamefont {Sakuma}}, \bibinfo {author} {\bibfnamefont
  {K.}~\bibnamefont {Tsukagoshi}}, \bibinfo {author} {\bibfnamefont
  {K.}~\bibnamefont {Suenaga}},\ and\ \bibinfo {author} {\bibfnamefont
  {T.}~\bibnamefont {Taniguchi}},\ }\bibfield  {title} {\bibinfo {title}
  {Tunable doping of rhenium and vanadium into transition metal dichalcogenides
  for two-dimensional electronics},\ }\href
  {https://doi.org/https://doi.org/10.1002/advs.202004438} {\bibfield
  {journal} {\bibinfo  {journal} {Advanced Science}\ }\textbf {\bibinfo
  {volume} {8}},\ \bibinfo {pages} {2004438} (\bibinfo {year}
  {2021})}\BibitemShut {NoStop}%
\bibitem [{\citenamefont {Wang}\ \emph {et~al.}(2021)\citenamefont {Wang},
  \citenamefont {Zhu}, \citenamefont {Seyler}, \citenamefont {Rivera},
  \citenamefont {Zheng}, \citenamefont {Wang}, \citenamefont {He},
  \citenamefont {Taniguchi}, \citenamefont {Watanabe}, \citenamefont {Yan},
  \citenamefont {Mandrus}, \citenamefont {Gamelin}, \citenamefont {Yao},\ and\
  \citenamefont {Xu}}]{Wang2021}%
  \BibitemOpen
  \bibfield  {author} {\bibinfo {author} {\bibfnamefont {X.}~\bibnamefont
  {Wang}}, \bibinfo {author} {\bibfnamefont {J.}~\bibnamefont {Zhu}}, \bibinfo
  {author} {\bibfnamefont {K.~L.}\ \bibnamefont {Seyler}}, \bibinfo {author}
  {\bibfnamefont {P.}~\bibnamefont {Rivera}}, \bibinfo {author} {\bibfnamefont
  {H.}~\bibnamefont {Zheng}}, \bibinfo {author} {\bibfnamefont
  {Y.}~\bibnamefont {Wang}}, \bibinfo {author} {\bibfnamefont {M.}~\bibnamefont
  {He}}, \bibinfo {author} {\bibfnamefont {T.}~\bibnamefont {Taniguchi}},
  \bibinfo {author} {\bibfnamefont {K.}~\bibnamefont {Watanabe}}, \bibinfo
  {author} {\bibfnamefont {J.}~\bibnamefont {Yan}}, \bibinfo {author}
  {\bibfnamefont {D.~G.}\ \bibnamefont {Mandrus}}, \bibinfo {author}
  {\bibfnamefont {D.~R.}\ \bibnamefont {Gamelin}}, \bibinfo {author}
  {\bibfnamefont {W.}~\bibnamefont {Yao}},\ and\ \bibinfo {author}
  {\bibfnamefont {X.}~\bibnamefont {Xu}},\ }\bibfield  {title} {\bibinfo
  {title} {Moire trions in mose2/wse2 heterobilayers},\ }\href
  {https://doi.org/10.1038/s41565-021-00969-2} {\bibfield  {journal} {\bibinfo
  {journal} {Nature Nanotechnology}\ }\textbf {\bibinfo {volume} {16}},\
  \bibinfo {pages} {1208} (\bibinfo {year} {2021})}\BibitemShut {NoStop}%
\bibitem [{\citenamefont {Brem}\ \emph {et~al.}(2020)\citenamefont {Brem},
  \citenamefont {Linderälv}, \citenamefont {Erhart},\ and\ \citenamefont
  {Malic}}]{Brem2020}%
  \BibitemOpen
  \bibfield  {author} {\bibinfo {author} {\bibfnamefont {S.}~\bibnamefont
  {Brem}}, \bibinfo {author} {\bibfnamefont {C.}~\bibnamefont {Linderälv}},
  \bibinfo {author} {\bibfnamefont {P.}~\bibnamefont {Erhart}},\ and\ \bibinfo
  {author} {\bibfnamefont {E.}~\bibnamefont {Malic}},\ }\bibfield  {title}
  {\bibinfo {title} {Tunable phases of moiré excitons in van der waals
  heterostructures},\ }\href {https://doi.org/10.1021/acs.nanolett.0c03019}
  {\bibfield  {journal} {\bibinfo  {journal} {Nano Letters}\ }\textbf {\bibinfo
  {volume} {20}},\ \bibinfo {pages} {8534} (\bibinfo {year}
  {2020})}\BibitemShut {NoStop}%
\bibitem [{\citenamefont {Regan}\ \emph {et~al.}(2020)\citenamefont {Regan},
  \citenamefont {Wang}, \citenamefont {Jin}, \citenamefont {Utama},
  \citenamefont {Gao}, \citenamefont {Wei}, \citenamefont {Zhao}, \citenamefont
  {Zhao}, \citenamefont {Zhang}, \citenamefont {Yumigeta}, \citenamefont
  {Blei}, \citenamefont {Carlstr\"{o}m}, \citenamefont {Watanabe},
  \citenamefont {Taniguchi}, \citenamefont {Tongay}, \citenamefont {Crommie},
  \citenamefont {Zettl},\ and\ \citenamefont {Wang}}]{Regan2020}%
  \BibitemOpen
  \bibfield  {author} {\bibinfo {author} {\bibfnamefont {E.~C.}\ \bibnamefont
  {Regan}}, \bibinfo {author} {\bibfnamefont {D.}~\bibnamefont {Wang}},
  \bibinfo {author} {\bibfnamefont {C.}~\bibnamefont {Jin}}, \bibinfo {author}
  {\bibfnamefont {M.~I.~B.}\ \bibnamefont {Utama}}, \bibinfo {author}
  {\bibfnamefont {B.}~\bibnamefont {Gao}}, \bibinfo {author} {\bibfnamefont
  {X.}~\bibnamefont {Wei}}, \bibinfo {author} {\bibfnamefont {S.}~\bibnamefont
  {Zhao}}, \bibinfo {author} {\bibfnamefont {W.}~\bibnamefont {Zhao}}, \bibinfo
  {author} {\bibfnamefont {Z.}~\bibnamefont {Zhang}}, \bibinfo {author}
  {\bibfnamefont {K.}~\bibnamefont {Yumigeta}}, \bibinfo {author}
  {\bibfnamefont {M.}~\bibnamefont {Blei}}, \bibinfo {author} {\bibfnamefont
  {J.~D.}\ \bibnamefont {Carlstr\"{o}m}}, \bibinfo {author} {\bibfnamefont
  {K.}~\bibnamefont {Watanabe}}, \bibinfo {author} {\bibfnamefont
  {T.}~\bibnamefont {Taniguchi}}, \bibinfo {author} {\bibfnamefont
  {S.}~\bibnamefont {Tongay}}, \bibinfo {author} {\bibfnamefont
  {M.}~\bibnamefont {Crommie}}, \bibinfo {author} {\bibfnamefont
  {A.}~\bibnamefont {Zettl}},\ and\ \bibinfo {author} {\bibfnamefont
  {F.}~\bibnamefont {Wang}},\ }\bibfield  {title} {\bibinfo {title} {Mott and
  generalized wigner crystal states in {WSe}2/{WS}2 moir{\'{e}}
  superlattices},\ }\href {https://doi.org/10.1038/s41586-020-2092-4}
  {\bibfield  {journal} {\bibinfo  {journal} {Nature}\ }\textbf {\bibinfo
  {volume} {579}},\ \bibinfo {pages} {359} (\bibinfo {year}
  {2020})}\BibitemShut {NoStop}%
\bibitem [{\citenamefont {Troue}\ \emph {et~al.}(2023)\citenamefont {Troue},
  \citenamefont {Figueiredo}, \citenamefont {Sigl}, \citenamefont {Paspalides},
  \citenamefont {Katzer}, \citenamefont {Taniguchi}, \citenamefont {Watanabe},
  \citenamefont {Selig}, \citenamefont {Knorr}, \citenamefont {Wurstbauer},\
  and\ \citenamefont {Holleitner}}]{Troue2023}%
  \BibitemOpen
  \bibfield  {author} {\bibinfo {author} {\bibfnamefont {M.}~\bibnamefont
  {Troue}}, \bibinfo {author} {\bibfnamefont {J.}~\bibnamefont {Figueiredo}},
  \bibinfo {author} {\bibfnamefont {L.}~\bibnamefont {Sigl}}, \bibinfo {author}
  {\bibfnamefont {C.}~\bibnamefont {Paspalides}}, \bibinfo {author}
  {\bibfnamefont {M.}~\bibnamefont {Katzer}}, \bibinfo {author} {\bibfnamefont
  {T.}~\bibnamefont {Taniguchi}}, \bibinfo {author} {\bibfnamefont
  {K.}~\bibnamefont {Watanabe}}, \bibinfo {author} {\bibfnamefont
  {M.}~\bibnamefont {Selig}}, \bibinfo {author} {\bibfnamefont
  {A.}~\bibnamefont {Knorr}}, \bibinfo {author} {\bibfnamefont
  {U.}~\bibnamefont {Wurstbauer}},\ and\ \bibinfo {author} {\bibfnamefont
  {A.~W.}\ \bibnamefont {Holleitner}},\ }\bibfield  {title} {\bibinfo {title}
  {Extended spatial coherence of interlayer excitons in
  ${\mathrm{mose}}_{2}/{\mathrm{wse}}_{2}$ heterobilayers},\ }\href
  {https://doi.org/10.1103/PhysRevLett.131.036902} {\bibfield  {journal}
  {\bibinfo  {journal} {Phys. Rev. Lett.}\ }\textbf {\bibinfo {volume} {131}},\
  \bibinfo {pages} {036902} (\bibinfo {year} {2023})}\BibitemShut {NoStop}%
\bibitem [{\citenamefont {Rosenberger}\ \emph {et~al.}(2020)\citenamefont
  {Rosenberger}, \citenamefont {Chuang}, \citenamefont {Phillips},
  \citenamefont {Oleshko}, \citenamefont {McCreary}, \citenamefont {Sivaram},
  \citenamefont {Hellberg},\ and\ \citenamefont {Jonker}}]{Rosenberger2020}%
  \BibitemOpen
  \bibfield  {author} {\bibinfo {author} {\bibfnamefont {M.~R.}\ \bibnamefont
  {Rosenberger}}, \bibinfo {author} {\bibfnamefont {H.-J.}\ \bibnamefont
  {Chuang}}, \bibinfo {author} {\bibfnamefont {M.}~\bibnamefont {Phillips}},
  \bibinfo {author} {\bibfnamefont {V.~P.}\ \bibnamefont {Oleshko}}, \bibinfo
  {author} {\bibfnamefont {K.~M.}\ \bibnamefont {McCreary}}, \bibinfo {author}
  {\bibfnamefont {S.~V.}\ \bibnamefont {Sivaram}}, \bibinfo {author}
  {\bibfnamefont {C.~S.}\ \bibnamefont {Hellberg}},\ and\ \bibinfo {author}
  {\bibfnamefont {B.~T.}\ \bibnamefont {Jonker}},\ }\bibfield  {title}
  {\bibinfo {title} {Twist angle-dependent atomic reconstruction and moire
  patterns in transition metal dichalcogenide heterostructures},\ }\href
  {https://doi.org/10.1021/acsnano.0c00088} {\bibfield  {journal} {\bibinfo
  {journal} {{ACS} Nano}\ }\textbf {\bibinfo {volume} {14}},\ \bibinfo {pages}
  {4550} (\bibinfo {year} {2020})}\BibitemShut {NoStop}%
\bibitem [{\citenamefont {Winkle}\ \emph {et~al.}(2023)\citenamefont {Winkle},
  \citenamefont {Craig}, \citenamefont {Carr}, \citenamefont {Dandu},
  \citenamefont {Bustillo}, \citenamefont {Ciston}, \citenamefont {Ophus},
  \citenamefont {Taniguchi}, \citenamefont {Watanabe}, \citenamefont {Raja},
  \citenamefont {Griffin},\ and\ \citenamefont {Bediako}}]{VanWinkle2023}%
  \BibitemOpen
  \bibfield  {author} {\bibinfo {author} {\bibfnamefont {M.~V.}\ \bibnamefont
  {Winkle}}, \bibinfo {author} {\bibfnamefont {I.~M.}\ \bibnamefont {Craig}},
  \bibinfo {author} {\bibfnamefont {S.}~\bibnamefont {Carr}}, \bibinfo {author}
  {\bibfnamefont {M.}~\bibnamefont {Dandu}}, \bibinfo {author} {\bibfnamefont
  {K.~C.}\ \bibnamefont {Bustillo}}, \bibinfo {author} {\bibfnamefont
  {J.}~\bibnamefont {Ciston}}, \bibinfo {author} {\bibfnamefont
  {C.}~\bibnamefont {Ophus}}, \bibinfo {author} {\bibfnamefont
  {T.}~\bibnamefont {Taniguchi}}, \bibinfo {author} {\bibfnamefont
  {K.}~\bibnamefont {Watanabe}}, \bibinfo {author} {\bibfnamefont
  {A.}~\bibnamefont {Raja}}, \bibinfo {author} {\bibfnamefont {S.~M.}\
  \bibnamefont {Griffin}},\ and\ \bibinfo {author} {\bibfnamefont {D.~K.}\
  \bibnamefont {Bediako}},\ }\bibfield  {title} {\bibinfo {title} {Rotational
  and dilational reconstruction in transition metal dichalcogenide moire
  bilayers},\ }\bibfield  {journal} {\bibinfo  {journal} {Nature
  Communications}\ }\textbf {\bibinfo {volume} {14}},\ \href
  {https://doi.org/10.1038/s41467-023-38504-7} {10.1038/s41467-023-38504-7}
  (\bibinfo {year} {2023})\BibitemShut {NoStop}%
\bibitem [{\citenamefont {Zhao}\ \emph {et~al.}(2023)\citenamefont {Zhao},
  \citenamefont {Li}, \citenamefont {Huang}, \citenamefont {Rupp},
  \citenamefont {G\"{o}ser}, \citenamefont {Vovk}, \citenamefont {Kruchinin},
  \citenamefont {Watanabe}, \citenamefont {Taniguchi}, \citenamefont {Bilgin},
  \citenamefont {Baimuratov},\ and\ \citenamefont
  {H\"{o}gele}}]{Zhao2023NatNano}%
  \BibitemOpen
  \bibfield  {author} {\bibinfo {author} {\bibfnamefont {S.}~\bibnamefont
  {Zhao}}, \bibinfo {author} {\bibfnamefont {Z.}~\bibnamefont {Li}}, \bibinfo
  {author} {\bibfnamefont {X.}~\bibnamefont {Huang}}, \bibinfo {author}
  {\bibfnamefont {A.}~\bibnamefont {Rupp}}, \bibinfo {author} {\bibfnamefont
  {J.}~\bibnamefont {G\"{o}ser}}, \bibinfo {author} {\bibfnamefont {I.~A.}\
  \bibnamefont {Vovk}}, \bibinfo {author} {\bibfnamefont {S.~Y.}\ \bibnamefont
  {Kruchinin}}, \bibinfo {author} {\bibfnamefont {K.}~\bibnamefont {Watanabe}},
  \bibinfo {author} {\bibfnamefont {T.}~\bibnamefont {Taniguchi}}, \bibinfo
  {author} {\bibfnamefont {I.}~\bibnamefont {Bilgin}}, \bibinfo {author}
  {\bibfnamefont {A.~S.}\ \bibnamefont {Baimuratov}},\ and\ \bibinfo {author}
  {\bibfnamefont {A.}~\bibnamefont {H\"{o}gele}},\ }\bibfield  {title}
  {\bibinfo {title} {Excitons in mesoscopically reconstructed moir{\'{e}}
  heterostructures},\ }\href {https://doi.org/10.1038/s41565-023-01356-9}
  {\bibfield  {journal} {\bibinfo  {journal} {Nature Nanotechnology}\ }\textbf
  {\bibinfo {volume} {18}},\ \bibinfo {pages} {572} (\bibinfo {year}
  {2023})}\BibitemShut {NoStop}%
\bibitem [{\citenamefont {Sung}\ \emph {et~al.}(2020)\citenamefont {Sung},
  \citenamefont {Zhou}, \citenamefont {Scuri}, \citenamefont {Z{\'{o}}lyomi},
  \citenamefont {Andersen}, \citenamefont {Yoo}, \citenamefont {Wild},
  \citenamefont {Joe}, \citenamefont {Gelly}, \citenamefont {Heo},
  \citenamefont {Magorrian}, \citenamefont {B{\'{e}}rub{\'{e}}}, \citenamefont
  {Valdivia}, \citenamefont {Taniguchi}, \citenamefont {Watanabe},
  \citenamefont {Lukin}, \citenamefont {Kim}, \citenamefont {Fal'ko},\ and\
  \citenamefont {Park}}]{Sung2020}%
  \BibitemOpen
  \bibfield  {author} {\bibinfo {author} {\bibfnamefont {J.}~\bibnamefont
  {Sung}}, \bibinfo {author} {\bibfnamefont {Y.}~\bibnamefont {Zhou}}, \bibinfo
  {author} {\bibfnamefont {G.}~\bibnamefont {Scuri}}, \bibinfo {author}
  {\bibfnamefont {V.}~\bibnamefont {Z{\'{o}}lyomi}}, \bibinfo {author}
  {\bibfnamefont {T.~I.}\ \bibnamefont {Andersen}}, \bibinfo {author}
  {\bibfnamefont {H.}~\bibnamefont {Yoo}}, \bibinfo {author} {\bibfnamefont
  {D.~S.}\ \bibnamefont {Wild}}, \bibinfo {author} {\bibfnamefont {A.~Y.}\
  \bibnamefont {Joe}}, \bibinfo {author} {\bibfnamefont {R.~J.}\ \bibnamefont
  {Gelly}}, \bibinfo {author} {\bibfnamefont {H.}~\bibnamefont {Heo}}, \bibinfo
  {author} {\bibfnamefont {S.~J.}\ \bibnamefont {Magorrian}}, \bibinfo {author}
  {\bibfnamefont {D.}~\bibnamefont {B{\'{e}}rub{\'{e}}}}, \bibinfo {author}
  {\bibfnamefont {A.~M.~M.}\ \bibnamefont {Valdivia}}, \bibinfo {author}
  {\bibfnamefont {T.}~\bibnamefont {Taniguchi}}, \bibinfo {author}
  {\bibfnamefont {K.}~\bibnamefont {Watanabe}}, \bibinfo {author}
  {\bibfnamefont {M.~D.}\ \bibnamefont {Lukin}}, \bibinfo {author}
  {\bibfnamefont {P.}~\bibnamefont {Kim}}, \bibinfo {author} {\bibfnamefont
  {V.~I.}\ \bibnamefont {Fal'ko}},\ and\ \bibinfo {author} {\bibfnamefont
  {H.}~\bibnamefont {Park}},\ }\bibfield  {title} {\bibinfo {title} {Broken
  mirror symmetry in excitonic response of reconstructed domains in twisted
  mose2/mose2 bilayers},\ }\href {https://doi.org/10.1038/s41565-020-0728-z}
  {\bibfield  {journal} {\bibinfo  {journal} {Nature Nanotechnology}\ }\textbf
  {\bibinfo {volume} {15}},\ \bibinfo {pages} {750} (\bibinfo {year}
  {2020})}\BibitemShut {NoStop}%
\bibitem [{\citenamefont {Parzefall}\ \emph {et~al.}(2021)\citenamefont
  {Parzefall}, \citenamefont {Holler}, \citenamefont {Scheuck}, \citenamefont
  {Beer}, \citenamefont {Lin}, \citenamefont {Peng}, \citenamefont {Monserrat},
  \citenamefont {Nagler}, \citenamefont {Kempf}, \citenamefont {Korn},\ and\
  \citenamefont {Schüller}}]{Parzefall2021}%
  \BibitemOpen
  \bibfield  {author} {\bibinfo {author} {\bibfnamefont {P.}~\bibnamefont
  {Parzefall}}, \bibinfo {author} {\bibfnamefont {J.}~\bibnamefont {Holler}},
  \bibinfo {author} {\bibfnamefont {M.}~\bibnamefont {Scheuck}}, \bibinfo
  {author} {\bibfnamefont {A.}~\bibnamefont {Beer}}, \bibinfo {author}
  {\bibfnamefont {K.-Q.}\ \bibnamefont {Lin}}, \bibinfo {author} {\bibfnamefont
  {B.}~\bibnamefont {Peng}}, \bibinfo {author} {\bibfnamefont {B.}~\bibnamefont
  {Monserrat}}, \bibinfo {author} {\bibfnamefont {P.}~\bibnamefont {Nagler}},
  \bibinfo {author} {\bibfnamefont {M.}~\bibnamefont {Kempf}}, \bibinfo
  {author} {\bibfnamefont {T.}~\bibnamefont {Korn}},\ and\ \bibinfo {author}
  {\bibfnamefont {C.}~\bibnamefont {Schüller}},\ }\bibfield  {title} {\bibinfo
  {title} {Moiré phonons in twisted mose2 –wse2 heterobilayers and their
  correlation with interlayer excitons},\ }\href
  {https://doi.org/10.1088/2053-1583/abf98e} {\bibfield  {journal} {\bibinfo
  {journal} {2D Materials}\ }\textbf {\bibinfo {volume} {8}},\ \bibinfo {pages}
  {035030} (\bibinfo {year} {2021})}\BibitemShut {NoStop}%
\bibitem [{\citenamefont {Andersen}\ \emph {et~al.}(2021)\citenamefont
  {Andersen}, \citenamefont {Scuri}, \citenamefont {Sushko}, \citenamefont
  {Greve}, \citenamefont {Sung}, \citenamefont {Zhou}, \citenamefont {Wild},
  \citenamefont {Gelly}, \citenamefont {Heo}, \citenamefont
  {B{\'{e}}rub{\'{e}}}, \citenamefont {Joe}, \citenamefont {Jauregui},
  \citenamefont {Watanabe}, \citenamefont {Taniguchi}, \citenamefont {Kim},
  \citenamefont {Park},\ and\ \citenamefont {Lukin}}]{Andersen2021}%
  \BibitemOpen
  \bibfield  {author} {\bibinfo {author} {\bibfnamefont {T.~I.}\ \bibnamefont
  {Andersen}}, \bibinfo {author} {\bibfnamefont {G.}~\bibnamefont {Scuri}},
  \bibinfo {author} {\bibfnamefont {A.}~\bibnamefont {Sushko}}, \bibinfo
  {author} {\bibfnamefont {K.~D.}\ \bibnamefont {Greve}}, \bibinfo {author}
  {\bibfnamefont {J.}~\bibnamefont {Sung}}, \bibinfo {author} {\bibfnamefont
  {Y.}~\bibnamefont {Zhou}}, \bibinfo {author} {\bibfnamefont {D.~S.}\
  \bibnamefont {Wild}}, \bibinfo {author} {\bibfnamefont {R.~J.}\ \bibnamefont
  {Gelly}}, \bibinfo {author} {\bibfnamefont {H.}~\bibnamefont {Heo}}, \bibinfo
  {author} {\bibfnamefont {D.}~\bibnamefont {B{\'{e}}rub{\'{e}}}}, \bibinfo
  {author} {\bibfnamefont {A.~Y.}\ \bibnamefont {Joe}}, \bibinfo {author}
  {\bibfnamefont {L.~A.}\ \bibnamefont {Jauregui}}, \bibinfo {author}
  {\bibfnamefont {K.}~\bibnamefont {Watanabe}}, \bibinfo {author}
  {\bibfnamefont {T.}~\bibnamefont {Taniguchi}}, \bibinfo {author}
  {\bibfnamefont {P.}~\bibnamefont {Kim}}, \bibinfo {author} {\bibfnamefont
  {H.}~\bibnamefont {Park}},\ and\ \bibinfo {author} {\bibfnamefont {M.~D.}\
  \bibnamefont {Lukin}},\ }\bibfield  {title} {\bibinfo {title} {Excitons in a
  reconstructed moire potential in twisted wse2/wse2 homobilayers},\ }\href
  {https://doi.org/10.1038/s41563-020-00873-5} {\bibfield  {journal} {\bibinfo
  {journal} {Nature Materials}\ }\textbf {\bibinfo {volume} {20}},\ \bibinfo
  {pages} {480} (\bibinfo {year} {2021})}\BibitemShut {NoStop}%
\bibitem [{\citenamefont {Rivera}\ \emph {et~al.}(2015)\citenamefont {Rivera},
  \citenamefont {Schaibley}, \citenamefont {Jones}, \citenamefont {Ross},
  \citenamefont {Wu}, \citenamefont {Aivazian}, \citenamefont {Klement},
  \citenamefont {Seyler}, \citenamefont {Clark}, \citenamefont {Ghimire},
  \citenamefont {Yan}, \citenamefont {Mandrus}, \citenamefont {Yao},\ and\
  \citenamefont {Xu}}]{Rivera2015}%
  \BibitemOpen
  \bibfield  {author} {\bibinfo {author} {\bibfnamefont {P.}~\bibnamefont
  {Rivera}}, \bibinfo {author} {\bibfnamefont {J.~R.}\ \bibnamefont
  {Schaibley}}, \bibinfo {author} {\bibfnamefont {A.~M.}\ \bibnamefont
  {Jones}}, \bibinfo {author} {\bibfnamefont {J.~S.}\ \bibnamefont {Ross}},
  \bibinfo {author} {\bibfnamefont {S.}~\bibnamefont {Wu}}, \bibinfo {author}
  {\bibfnamefont {G.}~\bibnamefont {Aivazian}}, \bibinfo {author}
  {\bibfnamefont {P.}~\bibnamefont {Klement}}, \bibinfo {author} {\bibfnamefont
  {K.}~\bibnamefont {Seyler}}, \bibinfo {author} {\bibfnamefont
  {G.}~\bibnamefont {Clark}}, \bibinfo {author} {\bibfnamefont {N.~J.}\
  \bibnamefont {Ghimire}}, \bibinfo {author} {\bibfnamefont {J.}~\bibnamefont
  {Yan}}, \bibinfo {author} {\bibfnamefont {D.~G.}\ \bibnamefont {Mandrus}},
  \bibinfo {author} {\bibfnamefont {W.}~\bibnamefont {Yao}},\ and\ \bibinfo
  {author} {\bibfnamefont {X.}~\bibnamefont {Xu}},\ }\bibfield  {title}
  {\bibinfo {title} {Observation of long-lived interlayer excitons in monolayer
  mose2–wse2 heterostructures},\ }\href {https://doi.org/10.1038/ncomms7242}
  {\bibfield  {journal} {\bibinfo  {journal} {Nature Communications}\ }\textbf
  {\bibinfo {volume} {6}},\ \bibinfo {pages} {6242} (\bibinfo {year}
  {2015})}\BibitemShut {NoStop}%
\bibitem [{\citenamefont {Nayak}\ \emph {et~al.}(2017)\citenamefont {Nayak},
  \citenamefont {Horbatenko}, \citenamefont {Ahn}, \citenamefont {Kim},
  \citenamefont {Lee}, \citenamefont {Ma}, \citenamefont {Jang}, \citenamefont
  {Lim}, \citenamefont {Kim}, \citenamefont {Ryu}, \citenamefont {Cheong},
  \citenamefont {Park},\ and\ \citenamefont {Shin}}]{Nayak2017}%
  \BibitemOpen
  \bibfield  {author} {\bibinfo {author} {\bibfnamefont {P.~K.}\ \bibnamefont
  {Nayak}}, \bibinfo {author} {\bibfnamefont {Y.}~\bibnamefont {Horbatenko}},
  \bibinfo {author} {\bibfnamefont {S.}~\bibnamefont {Ahn}}, \bibinfo {author}
  {\bibfnamefont {G.}~\bibnamefont {Kim}}, \bibinfo {author} {\bibfnamefont
  {J.-U.}\ \bibnamefont {Lee}}, \bibinfo {author} {\bibfnamefont {K.~Y.}\
  \bibnamefont {Ma}}, \bibinfo {author} {\bibfnamefont {A.-R.}\ \bibnamefont
  {Jang}}, \bibinfo {author} {\bibfnamefont {H.}~\bibnamefont {Lim}}, \bibinfo
  {author} {\bibfnamefont {D.}~\bibnamefont {Kim}}, \bibinfo {author}
  {\bibfnamefont {S.}~\bibnamefont {Ryu}}, \bibinfo {author} {\bibfnamefont
  {H.}~\bibnamefont {Cheong}}, \bibinfo {author} {\bibfnamefont
  {N.}~\bibnamefont {Park}},\ and\ \bibinfo {author} {\bibfnamefont {H.~S.}\
  \bibnamefont {Shin}},\ }\bibfield  {title} {\bibinfo {title} {Probing
  evolution of twist-angle-dependent interlayer excitons in mose 2 /wse 2 van
  der waals heterostructures (supporting information)},\ }\href
  {https://doi.org/10.1021/acsnano.7b00640} {\bibfield  {journal} {\bibinfo
  {journal} {ACS Nano}\ }\textbf {\bibinfo {volume} {11}},\ \bibinfo {pages}
  {4041} (\bibinfo {year} {2017})}\BibitemShut {NoStop}%
\bibitem [{\citenamefont {Nagler}\ \emph {et~al.}(2017)\citenamefont {Nagler},
  \citenamefont {Ballottin}, \citenamefont {Mitioglu}, \citenamefont
  {Mooshammer}, \citenamefont {Paradiso}, \citenamefont {Strunk}, \citenamefont
  {Huber}, \citenamefont {Chernikov}, \citenamefont {Christianen},
  \citenamefont {Sch\"{u}ller},\ and\ \citenamefont
  {Korn}}]{Nagler2017NatComm}%
  \BibitemOpen
  \bibfield  {author} {\bibinfo {author} {\bibfnamefont {P.}~\bibnamefont
  {Nagler}}, \bibinfo {author} {\bibfnamefont {M.~V.}\ \bibnamefont
  {Ballottin}}, \bibinfo {author} {\bibfnamefont {A.~A.}\ \bibnamefont
  {Mitioglu}}, \bibinfo {author} {\bibfnamefont {F.}~\bibnamefont
  {Mooshammer}}, \bibinfo {author} {\bibfnamefont {N.}~\bibnamefont
  {Paradiso}}, \bibinfo {author} {\bibfnamefont {C.}~\bibnamefont {Strunk}},
  \bibinfo {author} {\bibfnamefont {R.}~\bibnamefont {Huber}}, \bibinfo
  {author} {\bibfnamefont {A.}~\bibnamefont {Chernikov}}, \bibinfo {author}
  {\bibfnamefont {P.~C.~M.}\ \bibnamefont {Christianen}}, \bibinfo {author}
  {\bibfnamefont {C.}~\bibnamefont {Sch\"{u}ller}},\ and\ \bibinfo {author}
  {\bibfnamefont {T.}~\bibnamefont {Korn}},\ }\bibfield  {title} {\bibinfo
  {title} {Giant magnetic splitting inducing near-unity valley polarization in
  van der waals heterostructures},\ }\href
  {https://doi.org/10.1038/s41467-017-01748-1} {\bibfield  {journal} {\bibinfo
  {journal} {Nature Commun.}\ }\textbf {\bibinfo {volume} {8}},\ \bibinfo
  {pages} {1551} (\bibinfo {year} {2017})}\BibitemShut {NoStop}%
\bibitem [{\citenamefont {Gillen}\ and\ \citenamefont
  {Maultzsch}(2018)}]{Gillen2018}%
  \BibitemOpen
  \bibfield  {author} {\bibinfo {author} {\bibfnamefont {R.}~\bibnamefont
  {Gillen}}\ and\ \bibinfo {author} {\bibfnamefont {J.}~\bibnamefont
  {Maultzsch}},\ }\bibfield  {title} {\bibinfo {title} {Interlayer excitons in
  mose2/wse2 heterostructures from first principles},\ }\href
  {https://doi.org/10.1103/PhysRevB.97.165306} {\bibfield  {journal} {\bibinfo
  {journal} {Physical Review B}\ }\textbf {\bibinfo {volume} {97}},\ \bibinfo
  {pages} {165306} (\bibinfo {year} {2018})}\BibitemShut {NoStop}%
\bibitem [{\citenamefont {Zimmermann}\ \emph {et~al.}(2020)\citenamefont
  {Zimmermann}, \citenamefont {Kim}, \citenamefont {Hone}, \citenamefont
  {H\"{o}fer},\ and\ \citenamefont {Mette}}]{Zimmermann2020}%
  \BibitemOpen
  \bibfield  {author} {\bibinfo {author} {\bibfnamefont {J.~E.}\ \bibnamefont
  {Zimmermann}}, \bibinfo {author} {\bibfnamefont {Y.~D.}\ \bibnamefont {Kim}},
  \bibinfo {author} {\bibfnamefont {J.~C.}\ \bibnamefont {Hone}}, \bibinfo
  {author} {\bibfnamefont {U.}~\bibnamefont {H\"{o}fer}},\ and\ \bibinfo
  {author} {\bibfnamefont {G.}~\bibnamefont {Mette}},\ }\bibfield  {title}
  {\bibinfo {title} {Directional ultrafast charge transfer in a wse2/mose2
  heterostructure selectively probed by time-resolved {SHG} imaging
  microscopy},\ }\href {https://doi.org/10.1039/d0nh00396d} {\bibfield
  {journal} {\bibinfo  {journal} {Nanoscale Horizons}\ }\textbf {\bibinfo
  {volume} {5}},\ \bibinfo {pages} {1603} (\bibinfo {year} {2020})}\BibitemShut
  {NoStop}%
\bibitem [{\citenamefont {Chen}\ \emph {et~al.}(2022)\citenamefont {Chen},
  \citenamefont {Lian}, \citenamefont {Huang}, \citenamefont {Su},
  \citenamefont {Rashetnia}, \citenamefont {Yan}, \citenamefont {Blei},
  \citenamefont {Taniguchi}, \citenamefont {Watanabe}, \citenamefont {Tongay},
  \citenamefont {Wang}, \citenamefont {Zhang}, \citenamefont {Cui},\ and\
  \citenamefont {Shi}}]{Chen2022}%
  \BibitemOpen
  \bibfield  {author} {\bibinfo {author} {\bibfnamefont {D.}~\bibnamefont
  {Chen}}, \bibinfo {author} {\bibfnamefont {Z.}~\bibnamefont {Lian}}, \bibinfo
  {author} {\bibfnamefont {X.}~\bibnamefont {Huang}}, \bibinfo {author}
  {\bibfnamefont {Y.}~\bibnamefont {Su}}, \bibinfo {author} {\bibfnamefont
  {M.}~\bibnamefont {Rashetnia}}, \bibinfo {author} {\bibfnamefont
  {L.}~\bibnamefont {Yan}}, \bibinfo {author} {\bibfnamefont {M.}~\bibnamefont
  {Blei}}, \bibinfo {author} {\bibfnamefont {T.}~\bibnamefont {Taniguchi}},
  \bibinfo {author} {\bibfnamefont {K.}~\bibnamefont {Watanabe}}, \bibinfo
  {author} {\bibfnamefont {S.}~\bibnamefont {Tongay}}, \bibinfo {author}
  {\bibfnamefont {Z.}~\bibnamefont {Wang}}, \bibinfo {author} {\bibfnamefont
  {C.}~\bibnamefont {Zhang}}, \bibinfo {author} {\bibfnamefont {Y.-T.}\
  \bibnamefont {Cui}},\ and\ \bibinfo {author} {\bibfnamefont {S.-F.}\
  \bibnamefont {Shi}},\ }\bibfield  {title} {\bibinfo {title} {Tuning moiré
  excitons and correlated electronic states through layer degree of freedom},\
  }\href {https://doi.org/10.1038/s41467-022-32493-9} {\bibfield  {journal}
  {\bibinfo  {journal} {Nature Communications}\ }\textbf {\bibinfo {volume}
  {13}},\ \bibinfo {pages} {4810} (\bibinfo {year} {2022})}\BibitemShut
  {NoStop}%
\bibitem [{\citenamefont {F{\"o}rg}\ \emph {et~al.}(2021)\citenamefont
  {F{\"o}rg}, \citenamefont {Baimuratov}, \citenamefont {Kruchinin},
  \citenamefont {Vovk}, \citenamefont {Scherzer}, \citenamefont {F{\"o}rste},
  \citenamefont {Funk}, \citenamefont {Watanabe}, \citenamefont {Taniguchi},\
  and\ \citenamefont {H{\"o}gele}}]{Forg2021NatComm}%
  \BibitemOpen
  \bibfield  {author} {\bibinfo {author} {\bibfnamefont {M.}~\bibnamefont
  {F{\"o}rg}}, \bibinfo {author} {\bibfnamefont {A.~S.}\ \bibnamefont
  {Baimuratov}}, \bibinfo {author} {\bibfnamefont {S.~Y.}\ \bibnamefont
  {Kruchinin}}, \bibinfo {author} {\bibfnamefont {I.~A.}\ \bibnamefont {Vovk}},
  \bibinfo {author} {\bibfnamefont {J.}~\bibnamefont {Scherzer}}, \bibinfo
  {author} {\bibfnamefont {J.}~\bibnamefont {F{\"o}rste}}, \bibinfo {author}
  {\bibfnamefont {V.}~\bibnamefont {Funk}}, \bibinfo {author} {\bibfnamefont
  {K.}~\bibnamefont {Watanabe}}, \bibinfo {author} {\bibfnamefont
  {T.}~\bibnamefont {Taniguchi}},\ and\ \bibinfo {author} {\bibfnamefont
  {A.}~\bibnamefont {H{\"o}gele}},\ }\bibfield  {title} {\bibinfo {title}
  {Moiré excitons in mose$_2$-wse$_2$ heterobilayers and heterotrilayers},\
  }\href {https://doi.org/10.1038/s41467-021-21822-z} {\bibfield  {journal}
  {\bibinfo  {journal} {Nature Commun.}\ }\textbf {\bibinfo {volume} {12}},\
  \bibinfo {pages} {1656} (\bibinfo {year} {2021})}\BibitemShut {NoStop}%
\bibitem [{\citenamefont {Bai}\ \emph {et~al.}(2022)\citenamefont {Bai},
  \citenamefont {Li}, \citenamefont {Liu}, \citenamefont {Guo}, \citenamefont
  {Pack}, \citenamefont {Wang}, \citenamefont {Dean}, \citenamefont {Hone},\
  and\ \citenamefont {Zhu}}]{Bai2022}%
  \BibitemOpen
  \bibfield  {author} {\bibinfo {author} {\bibfnamefont {Y.}~\bibnamefont
  {Bai}}, \bibinfo {author} {\bibfnamefont {Y.}~\bibnamefont {Li}}, \bibinfo
  {author} {\bibfnamefont {S.}~\bibnamefont {Liu}}, \bibinfo {author}
  {\bibfnamefont {Y.}~\bibnamefont {Guo}}, \bibinfo {author} {\bibfnamefont
  {J.}~\bibnamefont {Pack}}, \bibinfo {author} {\bibfnamefont {J.}~\bibnamefont
  {Wang}}, \bibinfo {author} {\bibfnamefont {C.~R.}\ \bibnamefont {Dean}},
  \bibinfo {author} {\bibfnamefont {J.}~\bibnamefont {Hone}},\ and\ \bibinfo
  {author} {\bibfnamefont {X.~Y.}\ \bibnamefont {Zhu}},\ }\bibfield  {title}
  {\bibinfo {title} {Evidence for exciton crystals in a 2d semiconductor
  heterotrilayer},\ }\href@noop {} {\bibfield  {journal} {\bibinfo  {journal}
  {arXiv:2207.09601}\ } (\bibinfo {year} {2022})}\BibitemShut {NoStop}%
\bibitem [{\citenamefont {Slobodkin}\ \emph {et~al.}(2020)\citenamefont
  {Slobodkin}, \citenamefont {Mazuz-Harpaz}, \citenamefont {Refaely-Abramson},
  \citenamefont {Gazit}, \citenamefont {Steinberg},\ and\ \citenamefont
  {Rapaport}}]{Slobodkin2020}%
  \BibitemOpen
  \bibfield  {author} {\bibinfo {author} {\bibfnamefont {Y.}~\bibnamefont
  {Slobodkin}}, \bibinfo {author} {\bibfnamefont {Y.}~\bibnamefont
  {Mazuz-Harpaz}}, \bibinfo {author} {\bibfnamefont {S.}~\bibnamefont
  {Refaely-Abramson}}, \bibinfo {author} {\bibfnamefont {S.}~\bibnamefont
  {Gazit}}, \bibinfo {author} {\bibfnamefont {H.}~\bibnamefont {Steinberg}},\
  and\ \bibinfo {author} {\bibfnamefont {R.}~\bibnamefont {Rapaport}},\
  }\bibfield  {title} {\bibinfo {title} {Quantum phase transitions of trilayer
  excitons in atomically thin heterostructures},\ }\href
  {https://doi.org/10.1103/PhysRevLett.125.255301} {\bibfield  {journal}
  {\bibinfo  {journal} {Physical Review Letters}\ }\textbf {\bibinfo {volume}
  {125}},\ \bibinfo {pages} {255301} (\bibinfo {year} {2020})}\BibitemShut
  {NoStop}%
\bibitem [{\citenamefont {Lian}\ \emph {et~al.}(2023)\citenamefont {Lian},
  \citenamefont {Chen}, \citenamefont {Ma}, \citenamefont {Meng}, \citenamefont
  {Su}, \citenamefont {Yan}, \citenamefont {Huang}, \citenamefont {Wu},
  \citenamefont {Chen}, \citenamefont {Blei}, \citenamefont {Taniguchi},
  \citenamefont {Watanabe}, \citenamefont {Tongay}, \citenamefont {Zhang},
  \citenamefont {Cui},\ and\ \citenamefont {Shi}}]{Lian2023}%
  \BibitemOpen
  \bibfield  {author} {\bibinfo {author} {\bibfnamefont {Z.}~\bibnamefont
  {Lian}}, \bibinfo {author} {\bibfnamefont {D.}~\bibnamefont {Chen}}, \bibinfo
  {author} {\bibfnamefont {L.}~\bibnamefont {Ma}}, \bibinfo {author}
  {\bibfnamefont {Y.}~\bibnamefont {Meng}}, \bibinfo {author} {\bibfnamefont
  {Y.}~\bibnamefont {Su}}, \bibinfo {author} {\bibfnamefont {L.}~\bibnamefont
  {Yan}}, \bibinfo {author} {\bibfnamefont {X.}~\bibnamefont {Huang}}, \bibinfo
  {author} {\bibfnamefont {Q.}~\bibnamefont {Wu}}, \bibinfo {author}
  {\bibfnamefont {X.}~\bibnamefont {Chen}}, \bibinfo {author} {\bibfnamefont
  {M.}~\bibnamefont {Blei}}, \bibinfo {author} {\bibfnamefont {T.}~\bibnamefont
  {Taniguchi}}, \bibinfo {author} {\bibfnamefont {K.}~\bibnamefont {Watanabe}},
  \bibinfo {author} {\bibfnamefont {S.}~\bibnamefont {Tongay}}, \bibinfo
  {author} {\bibfnamefont {C.}~\bibnamefont {Zhang}}, \bibinfo {author}
  {\bibfnamefont {Y.-T.}\ \bibnamefont {Cui}},\ and\ \bibinfo {author}
  {\bibfnamefont {S.-F.}\ \bibnamefont {Shi}},\ }\bibfield  {title} {\bibinfo
  {title} {Quadrupolar excitons and hybridized interlayer mott insulator in a
  trilayer moiré superlattice},\ }\href
  {https://doi.org/10.1038/s41467-023-40288-9} {\bibfield  {journal} {\bibinfo
  {journal} {Nature Communications}\ }\textbf {\bibinfo {volume} {14}},\
  \bibinfo {pages} {4604} (\bibinfo {year} {2023})}\BibitemShut {NoStop}%
\bibitem [{\citenamefont {Malard}\ \emph
  {et~al.}(2013{\natexlab{a}})\citenamefont {Malard}, \citenamefont {Alencar},
  \citenamefont {Barboza}, \citenamefont {Mak},\ and\ \citenamefont
  {de~Paula}}]{Malard2013}%
  \BibitemOpen
  \bibfield  {author} {\bibinfo {author} {\bibfnamefont {L.~M.}\ \bibnamefont
  {Malard}}, \bibinfo {author} {\bibfnamefont {T.~V.}\ \bibnamefont {Alencar}},
  \bibinfo {author} {\bibfnamefont {A.~P.~M.}\ \bibnamefont {Barboza}},
  \bibinfo {author} {\bibfnamefont {K.~F.}\ \bibnamefont {Mak}},\ and\ \bibinfo
  {author} {\bibfnamefont {A.~M.}\ \bibnamefont {de~Paula}},\ }\bibfield
  {title} {\bibinfo {title} {Observation of intense second harmonic generation
  from mos$_2$ atomic crystals},\ }\href
  {https://doi.org/10.1103/PhysRevB.87.201401} {\bibfield  {journal} {\bibinfo
  {journal} {Physical Review B}\ }\textbf {\bibinfo {volume} {87}},\ \bibinfo
  {pages} {201401} (\bibinfo {year} {2013}{\natexlab{a}})}\BibitemShut
  {NoStop}%
\bibitem [{\citenamefont {Li}\ \emph {et~al.}(2020)\citenamefont {Li},
  \citenamefont {Lv}, \citenamefont {Ren}, \citenamefont {Li}, \citenamefont
  {Kong}, \citenamefont {Zeng}, \citenamefont {Tao}, \citenamefont {Wu},
  \citenamefont {Ma}, \citenamefont {Zhao}, \citenamefont {Wang}, \citenamefont
  {Dang}, \citenamefont {Chen}, \citenamefont {Liao}, \citenamefont {Duan},
  \citenamefont {Duan},\ and\ \citenamefont {Liu}}]{Li2020}%
  \BibitemOpen
  \bibfield  {author} {\bibinfo {author} {\bibfnamefont {Z.}~\bibnamefont
  {Li}}, \bibinfo {author} {\bibfnamefont {Y.}~\bibnamefont {Lv}}, \bibinfo
  {author} {\bibfnamefont {L.}~\bibnamefont {Ren}}, \bibinfo {author}
  {\bibfnamefont {J.}~\bibnamefont {Li}}, \bibinfo {author} {\bibfnamefont
  {L.}~\bibnamefont {Kong}}, \bibinfo {author} {\bibfnamefont {Y.}~\bibnamefont
  {Zeng}}, \bibinfo {author} {\bibfnamefont {Q.}~\bibnamefont {Tao}}, \bibinfo
  {author} {\bibfnamefont {R.}~\bibnamefont {Wu}}, \bibinfo {author}
  {\bibfnamefont {H.}~\bibnamefont {Ma}}, \bibinfo {author} {\bibfnamefont
  {B.}~\bibnamefont {Zhao}}, \bibinfo {author} {\bibfnamefont {D.}~\bibnamefont
  {Wang}}, \bibinfo {author} {\bibfnamefont {W.}~\bibnamefont {Dang}}, \bibinfo
  {author} {\bibfnamefont {K.}~\bibnamefont {Chen}}, \bibinfo {author}
  {\bibfnamefont {L.}~\bibnamefont {Liao}}, \bibinfo {author} {\bibfnamefont
  {X.}~\bibnamefont {Duan}}, \bibinfo {author} {\bibfnamefont {X.}~\bibnamefont
  {Duan}},\ and\ \bibinfo {author} {\bibfnamefont {Y.}~\bibnamefont {Liu}},\
  }\bibfield  {title} {\bibinfo {title} {Efficient strain modulation of 2d
  materials via polymer encapsulation},\ }\href
  {https://doi.org/10.1038/s41467-020-15023-3} {\bibfield  {journal} {\bibinfo
  {journal} {Nature Communications}\ }\textbf {\bibinfo {volume} {11}},\
  \bibinfo {pages} {1151} (\bibinfo {year} {2020})}\BibitemShut {NoStop}%
\bibitem [{\citenamefont {Hanbicki}\ \emph {et~al.}(2018)\citenamefont
  {Hanbicki}, \citenamefont {Chuang}, \citenamefont {Rosenberger},
  \citenamefont {Hellberg}, \citenamefont {Sivaram}, \citenamefont {McCreary},
  \citenamefont {Mazin},\ and\ \citenamefont {Jonker}}]{Hanbicki2018}%
  \BibitemOpen
  \bibfield  {author} {\bibinfo {author} {\bibfnamefont {A.~T.}\ \bibnamefont
  {Hanbicki}}, \bibinfo {author} {\bibfnamefont {H.~J.}\ \bibnamefont
  {Chuang}}, \bibinfo {author} {\bibfnamefont {M.~R.}\ \bibnamefont
  {Rosenberger}}, \bibinfo {author} {\bibfnamefont {C.~S.}\ \bibnamefont
  {Hellberg}}, \bibinfo {author} {\bibfnamefont {S.~V.}\ \bibnamefont
  {Sivaram}}, \bibinfo {author} {\bibfnamefont {K.~M.}\ \bibnamefont
  {McCreary}}, \bibinfo {author} {\bibfnamefont {I.~I.}\ \bibnamefont
  {Mazin}},\ and\ \bibinfo {author} {\bibfnamefont {B.~T.}\ \bibnamefont
  {Jonker}},\ }\bibfield  {title} {\bibinfo {title} {Double indirect interlayer
  exciton in a mose2/wse2 van der waals heterostructure},\ }\bibfield
  {journal} {\bibinfo  {journal} {ACS Nano}\ }\href
  {https://doi.org/10.1021/acsnano.8b01369} {10.1021/acsnano.8b01369} (\bibinfo
  {year} {2018})\BibitemShut {NoStop}%
\bibitem [{\citenamefont {Merkl}\ \emph {et~al.}(2020)\citenamefont {Merkl},
  \citenamefont {Mooshammer}, \citenamefont {Brem}, \citenamefont {Girnghuber},
  \citenamefont {Lin}, \citenamefont {Weigl}, \citenamefont {Liebich},
  \citenamefont {Yong}, \citenamefont {Gillen}, \citenamefont {Maultzsch},
  \citenamefont {Lupton}, \citenamefont {Malic},\ and\ \citenamefont
  {Huber}}]{Merkl2020}%
  \BibitemOpen
  \bibfield  {author} {\bibinfo {author} {\bibfnamefont {P.}~\bibnamefont
  {Merkl}}, \bibinfo {author} {\bibfnamefont {F.}~\bibnamefont {Mooshammer}},
  \bibinfo {author} {\bibfnamefont {S.}~\bibnamefont {Brem}}, \bibinfo {author}
  {\bibfnamefont {A.}~\bibnamefont {Girnghuber}}, \bibinfo {author}
  {\bibfnamefont {K.-Q.}\ \bibnamefont {Lin}}, \bibinfo {author} {\bibfnamefont
  {L.}~\bibnamefont {Weigl}}, \bibinfo {author} {\bibfnamefont
  {M.}~\bibnamefont {Liebich}}, \bibinfo {author} {\bibfnamefont {C.-K.}\
  \bibnamefont {Yong}}, \bibinfo {author} {\bibfnamefont {R.}~\bibnamefont
  {Gillen}}, \bibinfo {author} {\bibfnamefont {J.}~\bibnamefont {Maultzsch}},
  \bibinfo {author} {\bibfnamefont {J.~M.}\ \bibnamefont {Lupton}}, \bibinfo
  {author} {\bibfnamefont {E.}~\bibnamefont {Malic}},\ and\ \bibinfo {author}
  {\bibfnamefont {R.}~\bibnamefont {Huber}},\ }\bibfield  {title} {\bibinfo
  {title} {Twist-tailoring coulomb correlations in van der waals
  homobilayers},\ }\href {https://doi.org/10.1038/s41467-020-16069-z}
  {\bibfield  {journal} {\bibinfo  {journal} {Nature Communications}\ }\textbf
  {\bibinfo {volume} {11}},\ \bibinfo {pages} {2167} (\bibinfo {year}
  {2020})}\BibitemShut {NoStop}%
\bibitem [{\citenamefont {Lin}\ \emph {et~al.}(2021)\citenamefont {Lin},
  \citenamefont {Junior}, \citenamefont {Bauer}, \citenamefont {Peng},
  \citenamefont {Monserrat}, \citenamefont {Gmitra}, \citenamefont {Fabian},
  \citenamefont {Bange},\ and\ \citenamefont {Lupton}}]{Lin2021}%
  \BibitemOpen
  \bibfield  {author} {\bibinfo {author} {\bibfnamefont {K.-Q.}\ \bibnamefont
  {Lin}}, \bibinfo {author} {\bibfnamefont {P.~E.~F.}\ \bibnamefont {Junior}},
  \bibinfo {author} {\bibfnamefont {J.~M.}\ \bibnamefont {Bauer}}, \bibinfo
  {author} {\bibfnamefont {B.}~\bibnamefont {Peng}}, \bibinfo {author}
  {\bibfnamefont {B.}~\bibnamefont {Monserrat}}, \bibinfo {author}
  {\bibfnamefont {M.}~\bibnamefont {Gmitra}}, \bibinfo {author} {\bibfnamefont
  {J.}~\bibnamefont {Fabian}}, \bibinfo {author} {\bibfnamefont
  {S.}~\bibnamefont {Bange}},\ and\ \bibinfo {author} {\bibfnamefont {J.~M.}\
  \bibnamefont {Lupton}},\ }\bibfield  {title} {\bibinfo {title} {Twist-angle
  engineering of excitonic quantum interference and optical nonlinearities in
  stacked 2d semiconductors},\ }\href
  {https://doi.org/10.1038/s41467-021-21547-z} {\bibfield  {journal} {\bibinfo
  {journal} {Nature Communications}\ }\textbf {\bibinfo {volume} {12}},\
  \bibinfo {pages} {1553} (\bibinfo {year} {2021})}\BibitemShut {NoStop}%
\bibitem [{\citenamefont {Zhang}\ \emph {et~al.}(2020)\citenamefont {Zhang},
  \citenamefont {Zhang}, \citenamefont {Wu}, \citenamefont {Wang},
  \citenamefont {Gogna}, \citenamefont {Hou}, \citenamefont {Watanabe},
  \citenamefont {Taniguchi}, \citenamefont {Kulkarni}, \citenamefont {Kuo},
  \citenamefont {Forrest},\ and\ \citenamefont {Deng}}]{Zhang2020}%
  \BibitemOpen
  \bibfield  {author} {\bibinfo {author} {\bibfnamefont {L.}~\bibnamefont
  {Zhang}}, \bibinfo {author} {\bibfnamefont {Z.}~\bibnamefont {Zhang}},
  \bibinfo {author} {\bibfnamefont {F.}~\bibnamefont {Wu}}, \bibinfo {author}
  {\bibfnamefont {D.}~\bibnamefont {Wang}}, \bibinfo {author} {\bibfnamefont
  {R.}~\bibnamefont {Gogna}}, \bibinfo {author} {\bibfnamefont
  {S.}~\bibnamefont {Hou}}, \bibinfo {author} {\bibfnamefont {K.}~\bibnamefont
  {Watanabe}}, \bibinfo {author} {\bibfnamefont {T.}~\bibnamefont {Taniguchi}},
  \bibinfo {author} {\bibfnamefont {K.}~\bibnamefont {Kulkarni}}, \bibinfo
  {author} {\bibfnamefont {T.}~\bibnamefont {Kuo}}, \bibinfo {author}
  {\bibfnamefont {S.~R.}\ \bibnamefont {Forrest}},\ and\ \bibinfo {author}
  {\bibfnamefont {H.}~\bibnamefont {Deng}},\ }\bibfield  {title} {\bibinfo
  {title} {{Twist-angle dependence of moir{\'{e}} excitons in WS2/MoSe2
  heterobilayers}},\ }\href {https://doi.org/10.1038/s41467-020-19466-6}
  {\bibfield  {journal} {\bibinfo  {journal} {Nature Communications}\ }\textbf
  {\bibinfo {volume} {11}},\ \bibinfo {pages} {5888} (\bibinfo {year}
  {2020})}\BibitemShut {NoStop}%
\bibitem [{\citenamefont {Choi}\ \emph {et~al.}(2021)\citenamefont {Choi},
  \citenamefont {Florian}, \citenamefont {Steinhoff}, \citenamefont {Erben},
  \citenamefont {Tran}, \citenamefont {Kim}, \citenamefont {Sun}, \citenamefont
  {Quan}, \citenamefont {Claassen}, \citenamefont {Majumder}, \citenamefont
  {Hollingsworth}, \citenamefont {Taniguchi}, \citenamefont {Watanabe},
  \citenamefont {Ueno}, \citenamefont {Singh}, \citenamefont {Moody},
  \citenamefont {Jahnke},\ and\ \citenamefont {Li}}]{Choi2021}%
  \BibitemOpen
  \bibfield  {author} {\bibinfo {author} {\bibfnamefont {J.}~\bibnamefont
  {Choi}}, \bibinfo {author} {\bibfnamefont {M.}~\bibnamefont {Florian}},
  \bibinfo {author} {\bibfnamefont {A.}~\bibnamefont {Steinhoff}}, \bibinfo
  {author} {\bibfnamefont {D.}~\bibnamefont {Erben}}, \bibinfo {author}
  {\bibfnamefont {K.}~\bibnamefont {Tran}}, \bibinfo {author} {\bibfnamefont
  {D.~S.}\ \bibnamefont {Kim}}, \bibinfo {author} {\bibfnamefont
  {L.}~\bibnamefont {Sun}}, \bibinfo {author} {\bibfnamefont {J.}~\bibnamefont
  {Quan}}, \bibinfo {author} {\bibfnamefont {R.}~\bibnamefont {Claassen}},
  \bibinfo {author} {\bibfnamefont {S.}~\bibnamefont {Majumder}}, \bibinfo
  {author} {\bibfnamefont {J.~A.}\ \bibnamefont {Hollingsworth}}, \bibinfo
  {author} {\bibfnamefont {T.}~\bibnamefont {Taniguchi}}, \bibinfo {author}
  {\bibfnamefont {K.}~\bibnamefont {Watanabe}}, \bibinfo {author}
  {\bibfnamefont {K.}~\bibnamefont {Ueno}}, \bibinfo {author} {\bibfnamefont
  {A.}~\bibnamefont {Singh}}, \bibinfo {author} {\bibfnamefont
  {G.}~\bibnamefont {Moody}}, \bibinfo {author} {\bibfnamefont
  {F.}~\bibnamefont {Jahnke}},\ and\ \bibinfo {author} {\bibfnamefont
  {X.}~\bibnamefont {Li}},\ }\bibfield  {title} {\bibinfo {title} {Twist
  angle-dependent interlayer exciton lifetimes in van der waals
  heterostructures},\ }\href
  {https://doi.org/10.1103/PHYSREVLETT.126.047401/FIGURES/4/MEDIUM} {\bibfield
  {journal} {\bibinfo  {journal} {Physical Review Letters}\ }\textbf {\bibinfo
  {volume} {126}},\ \bibinfo {pages} {047401} (\bibinfo {year}
  {2021})}\BibitemShut {NoStop}%
\bibitem [{\citenamefont {Liu}\ \emph {et~al.}(2014)\citenamefont {Liu},
  \citenamefont {Zhang}, \citenamefont {Cao}, \citenamefont {Jin},
  \citenamefont {Qiu}, \citenamefont {Zhou}, \citenamefont {Zettl},
  \citenamefont {Yang}, \citenamefont {Louie},\ and\ \citenamefont
  {Wang}}]{liu2014evolution}%
  \BibitemOpen
  \bibfield  {author} {\bibinfo {author} {\bibfnamefont {K.}~\bibnamefont
  {Liu}}, \bibinfo {author} {\bibfnamefont {L.}~\bibnamefont {Zhang}}, \bibinfo
  {author} {\bibfnamefont {T.}~\bibnamefont {Cao}}, \bibinfo {author}
  {\bibfnamefont {C.}~\bibnamefont {Jin}}, \bibinfo {author} {\bibfnamefont
  {D.}~\bibnamefont {Qiu}}, \bibinfo {author} {\bibfnamefont {Q.}~\bibnamefont
  {Zhou}}, \bibinfo {author} {\bibfnamefont {A.}~\bibnamefont {Zettl}},
  \bibinfo {author} {\bibfnamefont {P.}~\bibnamefont {Yang}}, \bibinfo {author}
  {\bibfnamefont {S.~G.}\ \bibnamefont {Louie}},\ and\ \bibinfo {author}
  {\bibfnamefont {F.}~\bibnamefont {Wang}},\ }\bibfield  {title} {\bibinfo
  {title} {Evolution of interlayer coupling in twisted molybdenum disulfide
  bilayers},\ }\href@noop {} {\bibfield  {journal} {\bibinfo  {journal} {Nature
  communications}\ }\textbf {\bibinfo {volume} {5}},\ \bibinfo {pages} {4966}
  (\bibinfo {year} {2014})}\BibitemShut {NoStop}%
\bibitem [{\citenamefont {Zheng}\ \emph {et~al.}(2015)\citenamefont {Zheng},
  \citenamefont {Sun}, \citenamefont {Zhou}, \citenamefont {Liu}, \citenamefont
  {Liu}, \citenamefont {Shen},\ and\ \citenamefont {Fan}}]{zheng2015coupling}%
  \BibitemOpen
  \bibfield  {author} {\bibinfo {author} {\bibfnamefont {S.}~\bibnamefont
  {Zheng}}, \bibinfo {author} {\bibfnamefont {L.}~\bibnamefont {Sun}}, \bibinfo
  {author} {\bibfnamefont {X.}~\bibnamefont {Zhou}}, \bibinfo {author}
  {\bibfnamefont {F.}~\bibnamefont {Liu}}, \bibinfo {author} {\bibfnamefont
  {Z.}~\bibnamefont {Liu}}, \bibinfo {author} {\bibfnamefont {Z.}~\bibnamefont
  {Shen}},\ and\ \bibinfo {author} {\bibfnamefont {H.~J.}\ \bibnamefont
  {Fan}},\ }\bibfield  {title} {\bibinfo {title} {Coupling and interlayer
  exciton in twist-stacked ws2 bilayers},\ }\href@noop {} {\bibfield  {journal}
  {\bibinfo  {journal} {Advanced Optical Materials}\ }\textbf {\bibinfo
  {volume} {3}},\ \bibinfo {pages} {1600} (\bibinfo {year} {2015})}\BibitemShut
  {NoStop}%
\bibitem [{\citenamefont {Shi}\ \emph {et~al.}(2019)\citenamefont {Shi},
  \citenamefont {Li}, \citenamefont {Zhang}, \citenamefont {Feng},
  \citenamefont {Wang}, \citenamefont {Ren}, \citenamefont {Zhang},
  \citenamefont {Du}, \citenamefont {Wu}, \citenamefont {Sui} \emph
  {et~al.}}]{shi2019twisted}%
  \BibitemOpen
  \bibfield  {author} {\bibinfo {author} {\bibfnamefont {J.}~\bibnamefont
  {Shi}}, \bibinfo {author} {\bibfnamefont {Y.}~\bibnamefont {Li}}, \bibinfo
  {author} {\bibfnamefont {Z.}~\bibnamefont {Zhang}}, \bibinfo {author}
  {\bibfnamefont {W.}~\bibnamefont {Feng}}, \bibinfo {author} {\bibfnamefont
  {Q.}~\bibnamefont {Wang}}, \bibinfo {author} {\bibfnamefont {S.}~\bibnamefont
  {Ren}}, \bibinfo {author} {\bibfnamefont {J.}~\bibnamefont {Zhang}}, \bibinfo
  {author} {\bibfnamefont {W.}~\bibnamefont {Du}}, \bibinfo {author}
  {\bibfnamefont {X.}~\bibnamefont {Wu}}, \bibinfo {author} {\bibfnamefont
  {X.}~\bibnamefont {Sui}}, \emph {et~al.},\ }\bibfield  {title} {\bibinfo
  {title} {Twisted-angle-dependent optical behaviors of intralayer excitons and
  trions in ws2/wse2 heterostructure},\ }\href@noop {} {\bibfield  {journal}
  {\bibinfo  {journal} {Acs Photonics}\ }\textbf {\bibinfo {volume} {6}},\
  \bibinfo {pages} {3082} (\bibinfo {year} {2019})}\BibitemShut {NoStop}%
\bibitem [{\citenamefont {Volmer}\ \emph {et~al.}(2023)\citenamefont {Volmer},
  \citenamefont {Ersfeld}, \citenamefont {{Faria~Junior}}, \citenamefont
  {Waldecker}, \citenamefont {Parashar}, \citenamefont {Rathmann},
  \citenamefont {Dubey}, \citenamefont {Cojocariu}, \citenamefont {Feyer},
  \citenamefont {Watanabe}, \citenamefont {Taniguchi}, \citenamefont
  {Schneider}, \citenamefont {Plucinski}, \citenamefont {Stampfer},
  \citenamefont {Fabian},\ and\ \citenamefont {Beschoten}}]{Volmer2023npj}%
  \BibitemOpen
  \bibfield  {author} {\bibinfo {author} {\bibfnamefont {F.}~\bibnamefont
  {Volmer}}, \bibinfo {author} {\bibfnamefont {M.}~\bibnamefont {Ersfeld}},
  \bibinfo {author} {\bibfnamefont {P.~E.}\ \bibnamefont {{Faria~Junior}}},
  \bibinfo {author} {\bibfnamefont {L.}~\bibnamefont {Waldecker}}, \bibinfo
  {author} {\bibfnamefont {B.}~\bibnamefont {Parashar}}, \bibinfo {author}
  {\bibfnamefont {L.}~\bibnamefont {Rathmann}}, \bibinfo {author}
  {\bibfnamefont {S.}~\bibnamefont {Dubey}}, \bibinfo {author} {\bibfnamefont
  {I.}~\bibnamefont {Cojocariu}}, \bibinfo {author} {\bibfnamefont
  {V.}~\bibnamefont {Feyer}}, \bibinfo {author} {\bibfnamefont
  {K.}~\bibnamefont {Watanabe}}, \bibinfo {author} {\bibfnamefont
  {T.}~\bibnamefont {Taniguchi}}, \bibinfo {author} {\bibfnamefont {C.~M.}\
  \bibnamefont {Schneider}}, \bibinfo {author} {\bibfnamefont {L.}~\bibnamefont
  {Plucinski}}, \bibinfo {author} {\bibfnamefont {C.}~\bibnamefont {Stampfer}},
  \bibinfo {author} {\bibfnamefont {J.}~\bibnamefont {Fabian}},\ and\ \bibinfo
  {author} {\bibfnamefont {B.}~\bibnamefont {Beschoten}},\ }\bibfield  {title}
  {\bibinfo {title} {Twist angle dependent interlayer transfer of valley
  polarization from excitons to free charge carriers in {WSe}2/{MoSe}2
  heterobilayers},\ }\bibfield  {journal} {\bibinfo  {journal} {npj 2D
  Materials and Applications}\ }\textbf {\bibinfo {volume} {7}},\ \href
  {https://doi.org/10.1038/s41699-023-00420-1} {10.1038/s41699-023-00420-1}
  (\bibinfo {year} {2023})\BibitemShut {NoStop}%
\bibitem [{\citenamefont {Palekar}\ \emph {et~al.}(2023)\citenamefont
  {Palekar}, \citenamefont {Hagel}, \citenamefont {Rosa}, \citenamefont {Brem},
  \citenamefont {Shih}, \citenamefont {Limame}, \citenamefont {von Helversen},
  \citenamefont {Tongay}, \citenamefont {Malic},\ and\ \citenamefont
  {Reitzenstein}}]{palekar2023twist}%
  \BibitemOpen
  \bibfield  {author} {\bibinfo {author} {\bibfnamefont {C.~C.}\ \bibnamefont
  {Palekar}}, \bibinfo {author} {\bibfnamefont {J.}~\bibnamefont {Hagel}},
  \bibinfo {author} {\bibfnamefont {B.}~\bibnamefont {Rosa}}, \bibinfo {author}
  {\bibfnamefont {S.}~\bibnamefont {Brem}}, \bibinfo {author} {\bibfnamefont
  {C.-W.}\ \bibnamefont {Shih}}, \bibinfo {author} {\bibfnamefont
  {I.}~\bibnamefont {Limame}}, \bibinfo {author} {\bibfnamefont
  {M.}~\bibnamefont {von Helversen}}, \bibinfo {author} {\bibfnamefont
  {S.}~\bibnamefont {Tongay}}, \bibinfo {author} {\bibfnamefont
  {E.}~\bibnamefont {Malic}},\ and\ \bibinfo {author} {\bibfnamefont
  {S.}~\bibnamefont {Reitzenstein}},\ }\href@noop {} {\bibinfo {title} {Twist
  angle dependence of exciton resonances in wse$_2$/mose$_2$ moir\'e
  heterostructures}} (\bibinfo {year} {2023}),\ \Eprint
  {https://arxiv.org/abs/2309.16822} {arXiv:2309.16822 [cond-mat.mes-hall]}
  \BibitemShut {NoStop}%
\bibitem [{\citenamefont {Kunstmann}\ \emph {et~al.}(2018)\citenamefont
  {Kunstmann}, \citenamefont {Mooshammer}, \citenamefont {Nagler},
  \citenamefont {Chaves}, \citenamefont {Stein}, \citenamefont {Paradiso},
  \citenamefont {Plechinger}, \citenamefont {Strunk}, \citenamefont
  {Sch{\"u}ller}, \citenamefont {Seifert} \emph
  {et~al.}}]{kunstmann2018momentum}%
  \BibitemOpen
  \bibfield  {author} {\bibinfo {author} {\bibfnamefont {J.}~\bibnamefont
  {Kunstmann}}, \bibinfo {author} {\bibfnamefont {F.}~\bibnamefont
  {Mooshammer}}, \bibinfo {author} {\bibfnamefont {P.}~\bibnamefont {Nagler}},
  \bibinfo {author} {\bibfnamefont {A.}~\bibnamefont {Chaves}}, \bibinfo
  {author} {\bibfnamefont {F.}~\bibnamefont {Stein}}, \bibinfo {author}
  {\bibfnamefont {N.}~\bibnamefont {Paradiso}}, \bibinfo {author}
  {\bibfnamefont {G.}~\bibnamefont {Plechinger}}, \bibinfo {author}
  {\bibfnamefont {C.}~\bibnamefont {Strunk}}, \bibinfo {author} {\bibfnamefont
  {C.}~\bibnamefont {Sch{\"u}ller}}, \bibinfo {author} {\bibfnamefont
  {G.}~\bibnamefont {Seifert}}, \emph {et~al.},\ }\bibfield  {title} {\bibinfo
  {title} {Momentum-space indirect interlayer excitons in transition-metal
  dichalcogenide van der waals heterostructures},\ }\href@noop {} {\bibfield
  {journal} {\bibinfo  {journal} {Nature Physics}\ }\textbf {\bibinfo {volume}
  {14}},\ \bibinfo {pages} {801} (\bibinfo {year} {2018})}\BibitemShut
  {NoStop}%
\bibitem [{\citenamefont {Wo\ifmmode~\acute{z}\else \'{z}\fi{}niak}\ \emph
  {et~al.}(2020)\citenamefont {Wo\ifmmode~\acute{z}\else \'{z}\fi{}niak},
  \citenamefont {{Faria~Junior}}, \citenamefont {Seifert}, \citenamefont
  {Chaves},\ and\ \citenamefont {Kunstmann}}]{Wozniak2020PRB}%
  \BibitemOpen
  \bibfield  {author} {\bibinfo {author} {\bibfnamefont {T.}~\bibnamefont
  {Wo\ifmmode~\acute{z}\else \'{z}\fi{}niak}}, \bibinfo {author} {\bibfnamefont
  {P.~E.}\ \bibnamefont {{Faria~Junior}}}, \bibinfo {author} {\bibfnamefont
  {G.}~\bibnamefont {Seifert}}, \bibinfo {author} {\bibfnamefont
  {A.}~\bibnamefont {Chaves}},\ and\ \bibinfo {author} {\bibfnamefont
  {J.}~\bibnamefont {Kunstmann}},\ }\bibfield  {title} {\bibinfo {title}
  {Exciton $g$ factors of van der waals heterostructures from first-principles
  calculations},\ }\href {https://doi.org/10.1103/PhysRevB.101.235408}
  {\bibfield  {journal} {\bibinfo  {journal} {Phys. Rev. B}\ }\textbf {\bibinfo
  {volume} {101}},\ \bibinfo {pages} {235408} (\bibinfo {year}
  {2020})}\BibitemShut {NoStop}%
\bibitem [{\citenamefont {{Faria~Junior}}\ and\ \citenamefont
  {Fabian}(2023)}]{FariaJunior2023}%
  \BibitemOpen
  \bibfield  {author} {\bibinfo {author} {\bibfnamefont {P.~E.}\ \bibnamefont
  {{Faria~Junior}}}\ and\ \bibinfo {author} {\bibfnamefont {J.}~\bibnamefont
  {Fabian}},\ }\bibfield  {title} {\bibinfo {title} {Signatures of electric
  field and layer separation effects on the spin-valley physics of
  {MoSe}$_2$/{WSe}$_2$ heterobilayers: From energy bands to dipolar excitons},\
  }\href@noop {} {\bibfield  {journal} {\bibinfo  {journal} {Nanomaterials}\
  }\textbf {\bibinfo {volume} {13}},\ \bibinfo {pages} {1187} (\bibinfo {year}
  {2023})}\BibitemShut {NoStop}%
\bibitem [{\citenamefont {Jiang}\ \emph {et~al.}(2018)\citenamefont {Jiang},
  \citenamefont {Xu}, \citenamefont {Rasmita}, \citenamefont {Huang},
  \citenamefont {Li}, \citenamefont {Xiong},\ and\ \citenamefont
  {bo~Gao}}]{Jiang2018}%
  \BibitemOpen
  \bibfield  {author} {\bibinfo {author} {\bibfnamefont {C.}~\bibnamefont
  {Jiang}}, \bibinfo {author} {\bibfnamefont {W.}~\bibnamefont {Xu}}, \bibinfo
  {author} {\bibfnamefont {A.}~\bibnamefont {Rasmita}}, \bibinfo {author}
  {\bibfnamefont {Z.}~\bibnamefont {Huang}}, \bibinfo {author} {\bibfnamefont
  {K.}~\bibnamefont {Li}}, \bibinfo {author} {\bibfnamefont {Q.}~\bibnamefont
  {Xiong}},\ and\ \bibinfo {author} {\bibfnamefont {W.}~\bibnamefont
  {bo~Gao}},\ }\bibfield  {title} {\bibinfo {title} {Microsecond dark-exciton
  valley polarization memory in two-dimensional heterostructures},\ }\href
  {https://doi.org/10.1038/s41467-018-03174-3} {\bibfield  {journal} {\bibinfo
  {journal} {Nature Communications}\ }\textbf {\bibinfo {volume} {9}},\
  \bibinfo {pages} {753} (\bibinfo {year} {2018})}\BibitemShut {NoStop}%
\bibitem [{\citenamefont {Wang}\ \emph {et~al.}(2017)\citenamefont {Wang},
  \citenamefont {Robert}, \citenamefont {Glazov}, \citenamefont {Cadiz},
  \citenamefont {Courtade}, \citenamefont {Amand}, \citenamefont {Lagarde},
  \citenamefont {Taniguchi}, \citenamefont {Watanabe}, \citenamefont
  {Urbaszek},\ and\ \citenamefont {Marie}}]{Wang2017}%
  \BibitemOpen
  \bibfield  {author} {\bibinfo {author} {\bibfnamefont {G.}~\bibnamefont
  {Wang}}, \bibinfo {author} {\bibfnamefont {C.}~\bibnamefont {Robert}},
  \bibinfo {author} {\bibfnamefont {M.}~\bibnamefont {Glazov}}, \bibinfo
  {author} {\bibfnamefont {F.}~\bibnamefont {Cadiz}}, \bibinfo {author}
  {\bibfnamefont {E.}~\bibnamefont {Courtade}}, \bibinfo {author}
  {\bibfnamefont {T.}~\bibnamefont {Amand}}, \bibinfo {author} {\bibfnamefont
  {D.}~\bibnamefont {Lagarde}}, \bibinfo {author} {\bibfnamefont
  {T.}~\bibnamefont {Taniguchi}}, \bibinfo {author} {\bibfnamefont
  {K.}~\bibnamefont {Watanabe}}, \bibinfo {author} {\bibfnamefont
  {B.}~\bibnamefont {Urbaszek}},\ and\ \bibinfo {author} {\bibfnamefont
  {X.}~\bibnamefont {Marie}},\ }\bibfield  {title} {\bibinfo {title} {In-plane
  propagation of light in transition metal dichalcogenide monolayers: Optical
  selection rules},\ }\href {https://doi.org/10.1103/PhysRevLett.119.047401}
  {\bibfield  {journal} {\bibinfo  {journal} {Physical Review Letters}\
  }\textbf {\bibinfo {volume} {119}},\ \bibinfo {pages} {047401} (\bibinfo
  {year} {2017})}\BibitemShut {NoStop}%
\bibitem [{\citenamefont {Novoselov}\ \emph {et~al.}(2004)\citenamefont
  {Novoselov}, \citenamefont {Geim}, \citenamefont {Morozov}, \citenamefont
  {Jiang}, \citenamefont {Zhang}, \citenamefont {Dubonos}, \citenamefont
  {Grigorieva},\ and\ \citenamefont {Firsov}}]{Novoselov2004}%
  \BibitemOpen
  \bibfield  {author} {\bibinfo {author} {\bibfnamefont {K.~S.}\ \bibnamefont
  {Novoselov}}, \bibinfo {author} {\bibfnamefont {A.~K.}\ \bibnamefont {Geim}},
  \bibinfo {author} {\bibfnamefont {S.~V.}\ \bibnamefont {Morozov}}, \bibinfo
  {author} {\bibfnamefont {D.}~\bibnamefont {Jiang}}, \bibinfo {author}
  {\bibfnamefont {Y.}~\bibnamefont {Zhang}}, \bibinfo {author} {\bibfnamefont
  {S.~V.}\ \bibnamefont {Dubonos}}, \bibinfo {author} {\bibfnamefont {I.~V.}\
  \bibnamefont {Grigorieva}},\ and\ \bibinfo {author} {\bibfnamefont {A.~A.}\
  \bibnamefont {Firsov}},\ }\bibfield  {title} {\bibinfo {title} {Electric
  field effect in atomically thin carbon films},\ }\href
  {https://doi.org/10.1126/science.1102896} {\bibfield  {journal} {\bibinfo
  {journal} {Science}\ }\textbf {\bibinfo {volume} {306}},\ \bibinfo {pages}
  {666} (\bibinfo {year} {2004})}\BibitemShut {NoStop}%
\bibitem [{\citenamefont {Castellanos-Gomez}\ \emph {et~al.}(2014)\citenamefont
  {Castellanos-Gomez}, \citenamefont {Buscema}, \citenamefont {Molenaar},
  \citenamefont {Singh}, \citenamefont {Janssen}, \citenamefont {van~der
  Zant},\ and\ \citenamefont {Steele}}]{Castellanos-Gomez2014}%
  \BibitemOpen
  \bibfield  {author} {\bibinfo {author} {\bibfnamefont {A.}~\bibnamefont
  {Castellanos-Gomez}}, \bibinfo {author} {\bibfnamefont {M.}~\bibnamefont
  {Buscema}}, \bibinfo {author} {\bibfnamefont {R.}~\bibnamefont {Molenaar}},
  \bibinfo {author} {\bibfnamefont {V.}~\bibnamefont {Singh}}, \bibinfo
  {author} {\bibfnamefont {L.}~\bibnamefont {Janssen}}, \bibinfo {author}
  {\bibfnamefont {H.~S.~J.}\ \bibnamefont {van~der Zant}},\ and\ \bibinfo
  {author} {\bibfnamefont {G.~A.}\ \bibnamefont {Steele}},\ }\bibfield  {title}
  {\bibinfo {title} {Deterministic transfer of two-dimensional materials by
  all-dry viscoelastic stamping},\ }\href
  {https://doi.org/10.1088/2053-1583/1/1/011002} {\bibfield  {journal}
  {\bibinfo  {journal} {2D Materials}\ }\textbf {\bibinfo {volume} {1}},\
  \bibinfo {pages} {011002} (\bibinfo {year} {2014})}\BibitemShut {NoStop}%
\bibitem [{\citenamefont {Malard}\ \emph
  {et~al.}(2013{\natexlab{b}})\citenamefont {Malard}, \citenamefont {Alencar},
  \citenamefont {Barboza}, \citenamefont {Mak},\ and\ \citenamefont
  {de~Paula}}]{xubaka2013}%
  \BibitemOpen
  \bibfield  {author} {\bibinfo {author} {\bibfnamefont {L.~M.}\ \bibnamefont
  {Malard}}, \bibinfo {author} {\bibfnamefont {T.~V.}\ \bibnamefont {Alencar}},
  \bibinfo {author} {\bibfnamefont {A.~P.~M.}\ \bibnamefont {Barboza}},
  \bibinfo {author} {\bibfnamefont {K.~F.}\ \bibnamefont {Mak}},\ and\ \bibinfo
  {author} {\bibfnamefont {A.~M.}\ \bibnamefont {de~Paula}},\ }\bibfield
  {title} {\bibinfo {title} {{Observation of intense second harmonic generation
  from MoS$_{2}$ atomic crystals}},\ }\href
  {https://doi.org/10.1103/PhysRevB.87.201401} {\bibfield  {journal} {\bibinfo
  {journal} {Phys. Rev. B}\ }\textbf {\bibinfo {volume} {87}},\ \bibinfo
  {pages} {201401} (\bibinfo {year} {2013}{\natexlab{b}})}\BibitemShut
  {NoStop}%
\bibitem [{\citenamefont {Hsu}\ \emph {et~al.}(2014)\citenamefont {Hsu},
  \citenamefont {Zhao}, \citenamefont {Li}, \citenamefont {Chen}, \citenamefont
  {Chiu}, \citenamefont {Chang}, \citenamefont {Chou},\ and\ \citenamefont
  {Chang}}]{Hsu2014SecondHG}%
  \BibitemOpen
  \bibfield  {author} {\bibinfo {author} {\bibfnamefont {W.-T.}\ \bibnamefont
  {Hsu}}, \bibinfo {author} {\bibfnamefont {Z.}~\bibnamefont {Zhao}}, \bibinfo
  {author} {\bibfnamefont {L.}~\bibnamefont {Li}}, \bibinfo {author}
  {\bibfnamefont {C.-H.}\ \bibnamefont {Chen}}, \bibinfo {author}
  {\bibfnamefont {M.-H.}\ \bibnamefont {Chiu}}, \bibinfo {author}
  {\bibfnamefont {P.}~\bibnamefont {Chang}}, \bibinfo {author} {\bibfnamefont
  {Y.-C.}\ \bibnamefont {Chou}},\ and\ \bibinfo {author} {\bibfnamefont
  {W.}~\bibnamefont {Chang}},\ }\bibfield  {title} {\bibinfo {title} {Second
  harmonic generation from artificially stacked transition metal dichalcogenide
  twisted bilayers.},\ }\href
  {https://api.semanticscholar.org/CorpusID:17340877} {\bibfield  {journal}
  {\bibinfo  {journal} {ACS nano}\ }\textbf {\bibinfo {volume} {8 3}},\
  \bibinfo {pages} {2951} (\bibinfo {year} {2014})}\BibitemShut {NoStop}%
\bibitem [{\citenamefont {Blaha}\ \emph {et~al.}(2020)\citenamefont {Blaha},
  \citenamefont {Schwarz}, \citenamefont {Tran}, \citenamefont {Laskowski},
  \citenamefont {Madsen},\ and\ \citenamefont {Marks}}]{wien2k}%
  \BibitemOpen
  \bibfield  {author} {\bibinfo {author} {\bibfnamefont {P.}~\bibnamefont
  {Blaha}}, \bibinfo {author} {\bibfnamefont {K.}~\bibnamefont {Schwarz}},
  \bibinfo {author} {\bibfnamefont {F.}~\bibnamefont {Tran}}, \bibinfo {author}
  {\bibfnamefont {R.}~\bibnamefont {Laskowski}}, \bibinfo {author}
  {\bibfnamefont {G.~K.}\ \bibnamefont {Madsen}},\ and\ \bibinfo {author}
  {\bibfnamefont {L.~D.}\ \bibnamefont {Marks}},\ }\bibfield  {title} {\bibinfo
  {title} {Wien2k: An apw+lo program for calculating the properties of
  solids},\ }\href {https://doi.org/doi.org/10.1063/1.5143061} {\bibfield
  {journal} {\bibinfo  {journal} {The Journal of Chemical Physics}\ }\textbf
  {\bibinfo {volume} {152}},\ \bibinfo {pages} {074101} (\bibinfo {year}
  {2020})}\BibitemShut {NoStop}%
\bibitem [{\citenamefont {Perdew}\ \emph {et~al.}(1996)\citenamefont {Perdew},
  \citenamefont {Burke},\ and\ \citenamefont {Ernzerhof}}]{Perdew1996PRL}%
  \BibitemOpen
  \bibfield  {author} {\bibinfo {author} {\bibfnamefont {J.~P.}\ \bibnamefont
  {Perdew}}, \bibinfo {author} {\bibfnamefont {K.}~\bibnamefont {Burke}},\ and\
  \bibinfo {author} {\bibfnamefont {M.}~\bibnamefont {Ernzerhof}},\ }\bibfield
  {title} {\bibinfo {title} {Generalized gradient approximation made simple},\
  }\href {https://doi.org/10.1103/PhysRevLett.77.3865} {\bibfield  {journal}
  {\bibinfo  {journal} {Physical Review Letters}\ }\textbf {\bibinfo {volume}
  {77}},\ \bibinfo {pages} {3865} (\bibinfo {year} {1996})}\BibitemShut
  {NoStop}%
\bibitem [{\citenamefont {Grimme}\ \emph {et~al.}(2010)\citenamefont {Grimme},
  \citenamefont {Antony}, \citenamefont {Ehrlich},\ and\ \citenamefont
  {Krieg}}]{Grimme2010JCP}%
  \BibitemOpen
  \bibfield  {author} {\bibinfo {author} {\bibfnamefont {S.}~\bibnamefont
  {Grimme}}, \bibinfo {author} {\bibfnamefont {J.}~\bibnamefont {Antony}},
  \bibinfo {author} {\bibfnamefont {S.}~\bibnamefont {Ehrlich}},\ and\ \bibinfo
  {author} {\bibfnamefont {H.}~\bibnamefont {Krieg}},\ }\bibfield  {title}
  {\bibinfo {title} {{A consistent and accurate ab initio parametrization of
  density functional dispersion correction (DFT-D) for the 94 elements H-Pu}},\
  }\href {https://doi.org/10.1063/1.3382344} {\bibfield  {journal} {\bibinfo
  {journal} {The Journal of Chemical Physics}\ }\textbf {\bibinfo {volume}
  {132}},\ \bibinfo {pages} {154104} (\bibinfo {year} {2010})}\BibitemShut
  {NoStop}%
\bibitem [{\citenamefont {Singh}\ and\ \citenamefont
  {Nordstrom}(2006)}]{Singh2006}%
  \BibitemOpen
  \bibfield  {author} {\bibinfo {author} {\bibfnamefont {D.~J.}\ \bibnamefont
  {Singh}}\ and\ \bibinfo {author} {\bibfnamefont {L.}~\bibnamefont
  {Nordstrom}},\ }\href@noop {} {\emph {\bibinfo {title} {Planewaves,
  Pseudopotentials, and the LAPW method}}}\ (\bibinfo  {publisher} {Springer
  Science \& Business Media},\ \bibinfo {year} {2006})\BibitemShut {NoStop}%
\bibitem [{\citenamefont {Ambrosch-Draxl}\ and\ \citenamefont
  {Sofo}(2006)}]{Draxl2006CPC}%
  \BibitemOpen
  \bibfield  {author} {\bibinfo {author} {\bibfnamefont {C.}~\bibnamefont
  {Ambrosch-Draxl}}\ and\ \bibinfo {author} {\bibfnamefont {J.~O.}\
  \bibnamefont {Sofo}},\ }\bibfield  {title} {\bibinfo {title} {Linear optical
  properties of solids within the full-potential linearized augmented planewave
  method},\ }\href {https://doi.org/10.1016/j.cpc.2006.03.005} {\bibfield
  {journal} {\bibinfo  {journal} {Computer Physics Communications}\ }\textbf
  {\bibinfo {volume} {175}},\ \bibinfo {pages} {1} (\bibinfo {year}
  {2006})}\BibitemShut {NoStop}%
\bibitem [{\citenamefont {Rohlfing}\ and\ \citenamefont
  {Louie}(1998)}]{Rohlfing1998PRL}%
  \BibitemOpen
  \bibfield  {author} {\bibinfo {author} {\bibfnamefont {M.}~\bibnamefont
  {Rohlfing}}\ and\ \bibinfo {author} {\bibfnamefont {S.~G.}\ \bibnamefont
  {Louie}},\ }\bibfield  {title} {\bibinfo {title} {Electron-hole excitations
  in semiconductors and insulators},\ }\href
  {https://doi.org/10.1103/PhysRevLett.81.2312} {\bibfield  {journal} {\bibinfo
   {journal} {Physical Review Letters}\ }\textbf {\bibinfo {volume} {81}},\
  \bibinfo {pages} {2312} (\bibinfo {year} {1998})}\BibitemShut {NoStop}%
\bibitem [{\citenamefont {Rohlfing}\ and\ \citenamefont
  {Louie}(2000)}]{Rohlfing2000PRB}%
  \BibitemOpen
  \bibfield  {author} {\bibinfo {author} {\bibfnamefont {M.}~\bibnamefont
  {Rohlfing}}\ and\ \bibinfo {author} {\bibfnamefont {S.~G.}\ \bibnamefont
  {Louie}},\ }\bibfield  {title} {\bibinfo {title} {Electron-hole excitations
  and optical spectra from first principles},\ }\href
  {https://doi.org/10.1103/PhysRevB.62.4927} {\bibfield  {journal} {\bibinfo
  {journal} {Physical Review B}\ }\textbf {\bibinfo {volume} {62}},\ \bibinfo
  {pages} {4927} (\bibinfo {year} {2000})}\BibitemShut {NoStop}%
\bibitem [{\citenamefont {Ovesen}\ \emph {et~al.}(2019)\citenamefont {Ovesen},
  \citenamefont {Brem}, \citenamefont {Linder\"{a}lv}, \citenamefont {Kuisma},
  \citenamefont {Korn}, \citenamefont {Erhart}, \citenamefont {Selig},\ and\
  \citenamefont {Malic}}]{Ovesen2019CommPhys}%
  \BibitemOpen
  \bibfield  {author} {\bibinfo {author} {\bibfnamefont {S.}~\bibnamefont
  {Ovesen}}, \bibinfo {author} {\bibfnamefont {S.}~\bibnamefont {Brem}},
  \bibinfo {author} {\bibfnamefont {C.}~\bibnamefont {Linder\"{a}lv}}, \bibinfo
  {author} {\bibfnamefont {M.}~\bibnamefont {Kuisma}}, \bibinfo {author}
  {\bibfnamefont {T.}~\bibnamefont {Korn}}, \bibinfo {author} {\bibfnamefont
  {P.}~\bibnamefont {Erhart}}, \bibinfo {author} {\bibfnamefont
  {M.}~\bibnamefont {Selig}},\ and\ \bibinfo {author} {\bibfnamefont
  {E.}~\bibnamefont {Malic}},\ }\bibfield  {title} {\bibinfo {title}
  {Interlayer exciton dynamics in van der waals heterostructures},\ }\bibfield
  {journal} {\bibinfo  {journal} {Communications Physics}\ }\textbf {\bibinfo
  {volume} {2}},\ \href {https://doi.org/10.1038/s42005-019-0122-z}
  {10.1038/s42005-019-0122-z} (\bibinfo {year} {2019})\BibitemShut {NoStop}%
\bibitem [{\citenamefont {{Faria~Junior}}\ \emph {et~al.}(2023)\citenamefont
  {{Faria~Junior}}, \citenamefont {Naimer}, \citenamefont {McCreary},
  \citenamefont {Jonker}, \citenamefont {Finley}, \citenamefont {Crooker},
  \citenamefont {Fabian},\ and\ \citenamefont {Stier}}]{FariaJunior2023TDM}%
  \BibitemOpen
  \bibfield  {author} {\bibinfo {author} {\bibfnamefont {P.~E.}\ \bibnamefont
  {{Faria~Junior}}}, \bibinfo {author} {\bibfnamefont {T.}~\bibnamefont
  {Naimer}}, \bibinfo {author} {\bibfnamefont {K.~M.}\ \bibnamefont
  {McCreary}}, \bibinfo {author} {\bibfnamefont {B.~T.}\ \bibnamefont
  {Jonker}}, \bibinfo {author} {\bibfnamefont {J.~J.}\ \bibnamefont {Finley}},
  \bibinfo {author} {\bibfnamefont {S.~A.}\ \bibnamefont {Crooker}}, \bibinfo
  {author} {\bibfnamefont {J.}~\bibnamefont {Fabian}},\ and\ \bibinfo {author}
  {\bibfnamefont {A.~V.}\ \bibnamefont {Stier}},\ }\bibfield  {title} {\bibinfo
  {title} {Proximity-enhanced valley zeeman splitting at the {WS}$_2$/graphene
  interface},\ }\href {https://doi.org/10.1088/2053-1583/acd5df} {\bibfield
  {journal} {\bibinfo  {journal} {2D Materials}\ }\textbf {\bibinfo {volume}
  {10}},\ \bibinfo {pages} {034002} (\bibinfo {year} {2023})}\BibitemShut
  {NoStop}%
\bibitem [{\citenamefont {Laturia}\ \emph {et~al.}(2018)\citenamefont
  {Laturia}, \citenamefont {Van~de Put},\ and\ \citenamefont
  {Vandenberghe}}]{Laturia2018npj2D}%
  \BibitemOpen
  \bibfield  {author} {\bibinfo {author} {\bibfnamefont {A.}~\bibnamefont
  {Laturia}}, \bibinfo {author} {\bibfnamefont {M.~L.}\ \bibnamefont {Van~de
  Put}},\ and\ \bibinfo {author} {\bibfnamefont {W.~G.}\ \bibnamefont
  {Vandenberghe}},\ }\bibfield  {title} {\bibinfo {title} {Dielectric
  properties of hexagonal boron nitride and transition metal dichalcogenides:
  from monolayer to bulk},\ }\href@noop {} {\bibfield  {journal} {\bibinfo
  {journal} {npj 2D Materials and Applications}\ }\textbf {\bibinfo {volume}
  {2}},\ \bibinfo {pages} {6} (\bibinfo {year} {2018})}\BibitemShut {NoStop}%
\bibitem [{\citenamefont {Berkelbach}\ \emph {et~al.}(2013)\citenamefont
  {Berkelbach}, \citenamefont {Hybertsen},\ and\ \citenamefont
  {Reichman}}]{Berkelbach2013PRB}%
  \BibitemOpen
  \bibfield  {author} {\bibinfo {author} {\bibfnamefont {T.~C.}\ \bibnamefont
  {Berkelbach}}, \bibinfo {author} {\bibfnamefont {M.~S.}\ \bibnamefont
  {Hybertsen}},\ and\ \bibinfo {author} {\bibfnamefont {D.~R.}\ \bibnamefont
  {Reichman}},\ }\bibfield  {title} {\bibinfo {title} {Theory of neutral and
  charged excitons in monolayer transition metal dichalcogenides},\ }\href
  {https://doi.org/10.1103/PhysRevB.88.045318} {\bibfield  {journal} {\bibinfo
  {journal} {Phys. Rev. B}\ }\textbf {\bibinfo {volume} {88}},\ \bibinfo
  {pages} {045318} (\bibinfo {year} {2013})}\BibitemShut {NoStop}%
\bibitem [{\citenamefont {Stier}\ \emph {et~al.}(2018)\citenamefont {Stier},
  \citenamefont {Wilson}, \citenamefont {Velizhanin}, \citenamefont {Kono},
  \citenamefont {Xu},\ and\ \citenamefont {Crooker}}]{Stier2018PRL}%
  \BibitemOpen
  \bibfield  {author} {\bibinfo {author} {\bibfnamefont {A.~V.}\ \bibnamefont
  {Stier}}, \bibinfo {author} {\bibfnamefont {N.~P.}\ \bibnamefont {Wilson}},
  \bibinfo {author} {\bibfnamefont {K.~A.}\ \bibnamefont {Velizhanin}},
  \bibinfo {author} {\bibfnamefont {J.}~\bibnamefont {Kono}}, \bibinfo {author}
  {\bibfnamefont {X.}~\bibnamefont {Xu}},\ and\ \bibinfo {author}
  {\bibfnamefont {S.~A.}\ \bibnamefont {Crooker}},\ }\bibfield  {title}
  {\bibinfo {title} {Magnetooptics of exciton rydberg states in a monolayer
  semiconductor},\ }\href {https://doi.org/10.1103/PhysRevLett.120.057405}
  {\bibfield  {journal} {\bibinfo  {journal} {Phys. Rev. Lett.}\ }\textbf
  {\bibinfo {volume} {120}},\ \bibinfo {pages} {057405} (\bibinfo {year}
  {2018})}\BibitemShut {NoStop}%
\end{thebibliography}
\end{document}


\preprint{APS/123-QED}

\title{Supplementary Information \\ 
Amplification of interlayer exciton emission in \\  twisted WSe$_2$/WSe$_2$/MoSe$_2$ heterotrilayers}

\author{Chirag C. Palekar}\email{c.palekar@tu-berlin.de}
\affiliation{Institute of Solid State Physics, Technische Universität Berlin, 10623 Berlin, Germany}

\author{Paulo E. {Faria~Junior}} \email{fariajunior.pe@gmail.com} 
\affiliation{Institute of Theoretical Physics, University of Regensburg, 93040 Regensburg, Germany}

\author{Barbara Rosa}\email{rosa@physik.tu-berlin.de} 
\affiliation{Institute of Solid State Physics, Technische Universität Berlin, 10623 Berlin, Germany}

\author{Frederico B. Sousa}
\affiliation{Departamento de Física, Universidade Federal de Minas Gerais, Belo Horizonte, Minas Gerais 30123-970}

\author{Leandro M. Malard}
\affiliation{Departamento de Física, Universidade Federal de Minas Gerais, Belo Horizonte, Minas Gerais 30123-970}
\author{Jaroslav Fabian}
\affiliation{Institute of Theoretical Physics, University of Regensburg, 93040 Regensburg, Germany}

\author{Stephan Reitzenstein}
\affiliation{Institute of Solid State Physics, Technische Universität Berlin, 10623 Berlin, Germany} 

\maketitle

\tableofcontents

\setcounter{table}{0}
\renewcommand{\tablename}{Table S}
\renewcommand{\thetable}{\arabic{table}}

\setcounter{figure}{0}
\renewcommand{\figurename}{Figure S}
\renewcommand{\thefigure}{\arabic{figure}}


\clearpage

\newpage

\section{Twist angle determination based on SHG measurements}

The twist angle between the monolayer of the heterostructure was determined by polarization resolved second harmonic generation (SHG) measurements. Fig. S1 shows SHG measurements for S1 and S2 along with the extracted twist angle in HBL and HTL regions. The laser excitation used for SHG measurement was at 1313 nm and SHG signal was measured at 656 nm. The intensity as function of excitation laser polarization is measured from the constituent MLs, which were illuminated with a linearly polarized light of a femtosecond mode-locked laser. In response, high intensity maxima are observed in six-fold symmetry, as each maximum indicates the armchair direction on hexagonal crystal lattice of TMDC ML. Comparing the SHG response from the constituent MLs of the HBL and HTL, the twist angle can be determined. To distinguish between R or H-type stacking, we recorded the SHG response from the HBL and HTL regions. Here, a reduction in intensity of SHG signal from the heterostructure with respect to ML indicates the H-type stacking. 

\begin{figure}[h!]
\centering
\includegraphics[width=\textwidth]{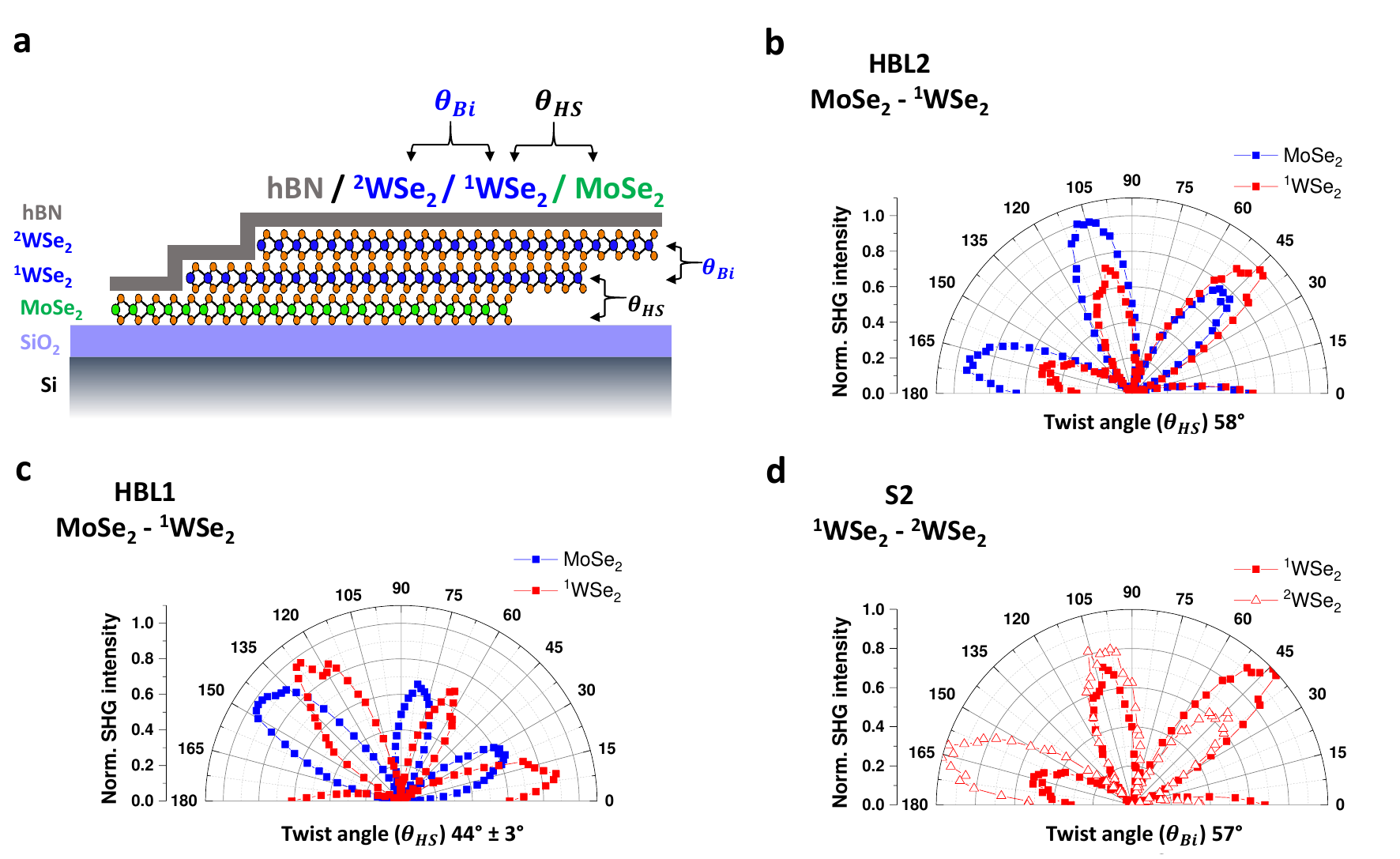}  
\caption{\textbf{ Twist angle determination using SHG.} (\textbf{a}) Schematic representation of the HBL and HTL system used for the investigation. (\textbf{b,c,d}) SHG signal from HBL and HTL regions of S1 and S2 with different twist angles as function of excitation laser polarization angle. Correspondingly, the polarization resolved SHG signal is measured to identify the crystallographic directions. The SHG response is then fitted with a Gaussian function to extract the position of each intensity maxima. The twist angle then determine based on difference between the positions of intensity maxima from each ML. See Methods Section C in the main text for further details.}
\label{Sulp. fig1}
\end{figure}
\clearpage

\newpage
\section{PLE resonances of HBL and HTL regions}

\begin{figure}[h!]
 \centering
 \includegraphics[width=\textwidth]{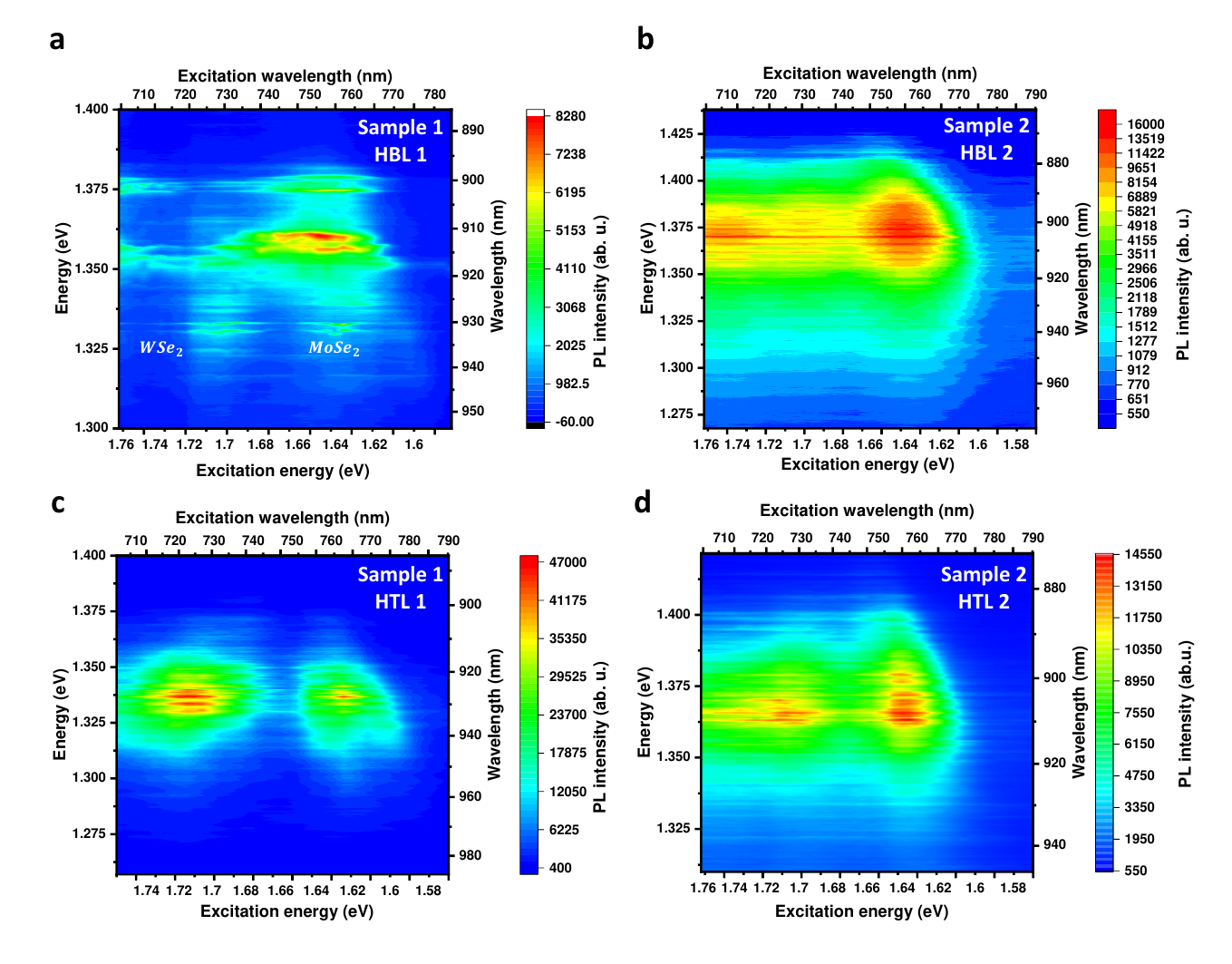}
 \caption{ \textbf{False color PLE map of HBL and HTL regions from S1 and S2.} Interlayer exciton emission as a function of excitation energy from S1 (\textbf{a,c}) and S2 (\textbf{b,d}). See Methods Section B in the main text for further details.}
 \label{Sulp. fig2}
\end{figure}

\begin{table*}[h!]
    \begin{tabular}{|p{3cm}|p{3cm}|p{3cm}|p{3cm}|p{3cm}|}
\hline
 {} Region (sample)& \multicolumn{2}{c}{HBL(S1)} \vline & \multicolumn{2}{c}{HTL(S1)} \vline\\
   \hline
    {}   Material & WSe$_2$  & MoSe$_2$  & WSe$_2$ & MoSe$_2$ \\
    \hline
  Energy (eV)    & 1.727  & 1.639  & 1.719 & 1.625 \\
  FWHM (meV)  & 53.44  & 47.87  & 63.56 & 36.98  \\
  Area   & 5854 & 5531  & 9389 & 4045  \\
  \hline
  {}  Region (sample) & \multicolumn{2}{c}{HBL(S2)} \vline& \multicolumn{2}{c}{HTL(S2)}\vline \\
   \hline
    {}   Material  & WSe$_2$  & MoSe$_2$  & WSe$_2$ & MoSe$_2$ \\
    \hline
  Energy (eV)    & 1.730  & 1.640  & 1.742 & 1.640 \\
  FWHM (meV)  & 28.38  & 36.86  & 35.9 & 47.36  \\
  Area   & 1028 & 4171  & 1856 & 5668  \\
  \hline 
\end{tabular}
\caption{ {\textbf {PLE resonances.}} Energy and full width half maximum (FWHM) extracted by fitting double Gaussian function to the PLE response demonstrated in Fig.2(b,c) from HBL and HTL regions of S1 and S2.}
\label{Sulp. table1}
\end{table*}

\clearpage

\newpage

\section{Spatial homogeneity of PL enhancement }
\begin{figure}[h!]
 \centering
 \includegraphics[width=\textwidth]{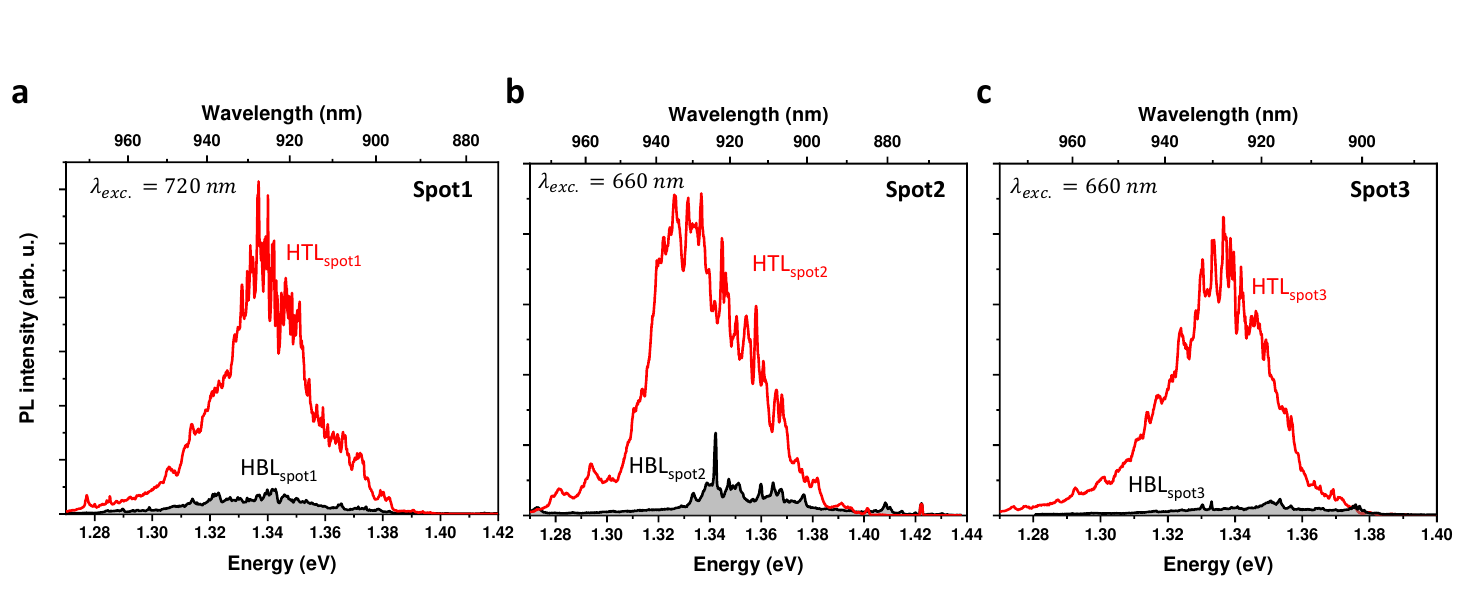}
 \caption{ \textbf{iX emission collected at different spots from the HBL1 and HTL1 regions.} (\textbf{a,b,c}) show the disparity between emission intensity of HBL and HTL region. For reasonable comparison of intensity, the excitation wavelength ($\lambda$$_{exc.}$), power and exposure times are kept identical for while recording the response for HBL and HTL regions shown in the same panel. }
 \label{Sulp. fig3}
\end{figure}

\begin{figure}[h!]
 \centering
 \includegraphics[width=0.6\textwidth]{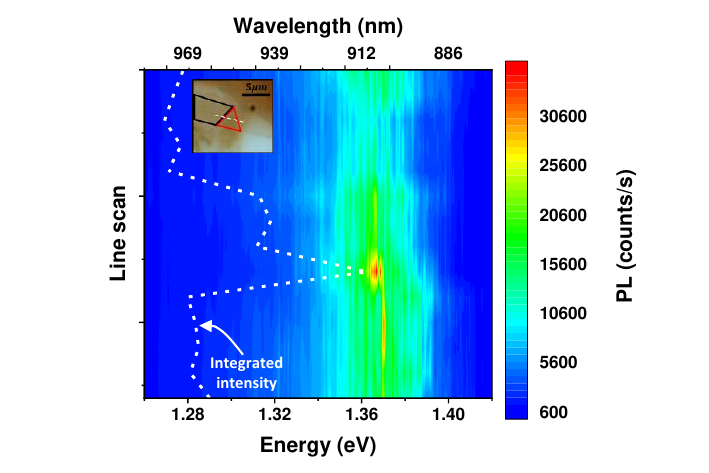}
 \caption{ \textbf{False color line scan PL map of sample S2,}  presenting the iX emission from HBL and HTL regions. The white dotted line in PL map manifests the integrated intensity profile, showcasing enhancement of emission from the HTL2 region when compared to HBL2 region. }
 \label{Sulp. fig4}
\end{figure}

\clearpage

\newpage

\section{Density functional theory calculations}

\begin{figure}[h!]
 \centering
 \includegraphics{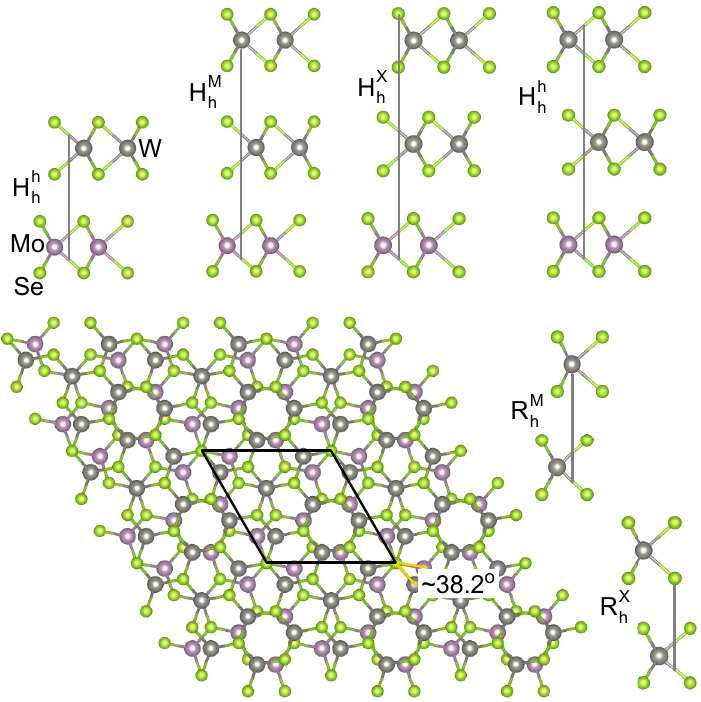}
 \caption{Crystal structures considered for the studied heterobilayers (HBL) and heterotrilayers (HTL).}
 \label{fig:structures}
\end{figure}

\renewcommand{\arraystretch}{1.5} 

\begin{table}[h!]
\caption{Structural and energetic parameters of the calculated structures. Energy/atom given in Ry, thickness d given in \textrm{\AA}, and interlayer distance $i$ given in \textrm{\AA}. The trilayer structures identified with $*$ indicate the lowest energy. The valence band splitting, $\Delta_v$ in meV, indicates the energy separation at the K point between $^1$WSe$_2$ and $^2$WSe$_2$.}
\begin{centering}
\begin{tabular}{cccccccc}
\hline
\hline
 & H$^{\textrm{h}}_{\textrm{h}}$ & 
H$^{\textrm{h}}_{\textrm{h}}$/H$^{\textrm{M}}_{\textrm{h}}$ & 
H$^{\textrm{h}}_{\textrm{h}}$/H$^{\textrm{X}}_{\textrm{h}}$ &
H$^{\textrm{h}}_{\textrm{h}}$/H$^{\textrm{h}*}_{\textrm{h}}$ &
$38.2 \degree$ & 
$38.2 \degree$/R$^{\textrm{X}*}_{\textrm{h}}$ &
$38.2 \degree$/R$^{\textrm{M}}_{\textrm{h}}$\tabularnewline
\hline
Ene./atom & $-$9978.961221 & $-$11325.243173 & $-$11325.243637 & $-$11325.243886 & $-$9978.960588 & $-$11325.243416 & $-$11325.243411\tabularnewline
\hline
d{[}Mo{]}  & 3.3607 & 3.3615 & 3.3611 & 3.3612 & 3.3576 & 3.3636 & 3.3597\tabularnewline
d{[}1W{]} & 3.3872 & 3.3879 & 3.3855 & 3.3874 & 3.3838 & 3.3872 & 3.3838\tabularnewline
d{[}2W{]} & --- & 3.3885 & 3.3872 & 3.3881 & --- & 3.3871 & 3.3859\tabularnewline
\hline
i{[}Mo$-$1W{]} & 3.0636 & 3.0700 & 3.0233 & 3.0629 & 3.3260 & 3.3853 & 3.3892\tabularnewline
i{[}1W$-$2W{]} & --- & 3.6602 & 3.1539 & 3.0505 & --- & 3.0561 & 3.0281\tabularnewline
\hline
$\Delta_v$ & --- & 25.7 & 19.2 & 6.0 & --- & 87.3 & -65.3 \tabularnewline
\hline
\hline
\end{tabular}
\end{centering}
\label{tab:structure}
\end{table}

\begin{figure}[h!]
 \centering
 \includegraphics{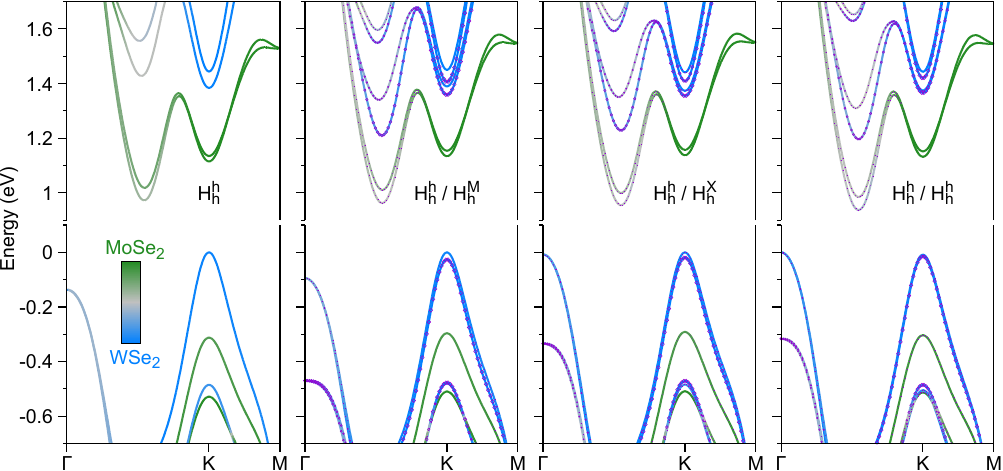}
 \caption{Calculated band structures for HBL and HTL systems with H-type stacking ($60 \degree$). The color code represents the MoSe$_2$ and WSe$_2$ layers. For HTL systems, the second WSe$_2$ layer is identified by the circles.}
 \label{fig:Hhh}
\end{figure}

\begin{figure}[h!]
 \centering
 \includegraphics{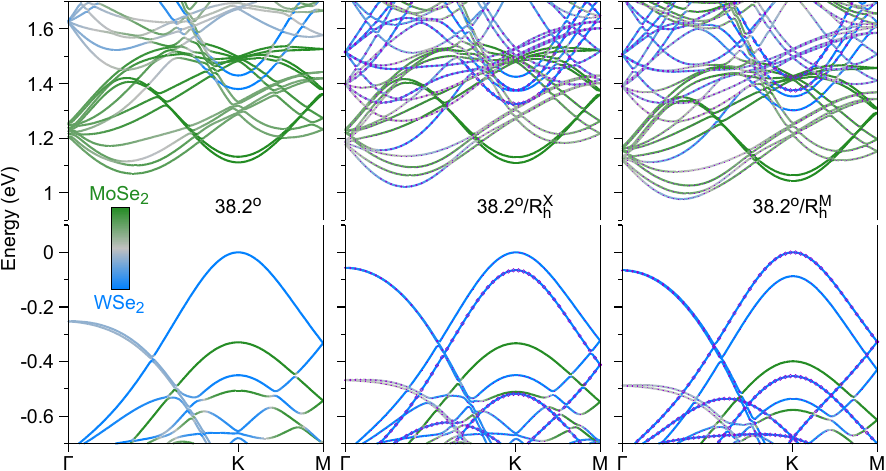}
 \caption{Same as in Fig.S~\ref{fig:Hhh} but for the structures with $38.2 \degree$ twist angle between MoSe$_2$ and WSe$_2$, and R-type stacking between the two WSe$_2$ layers.}
 \label{fig:38.2}
\end{figure}

\setlength{\tabcolsep}{6pt} 
\renewcommand{\arraystretch}{1.5} 

\begin{table}[h!]
\caption{Calculated transition energies and momentum matrix elements for the $60 \degree$ systems. The momentum matrix elements are calculated as $\left|p_{vc}^{\alpha}\right| = \frac{\hbar}{m_0} \left\langle v, K \left| \vec{p} \cdot \hat{\alpha} \right| c, K \right\rangle$ with $\alpha = +,-,z$ polarization of light. Energies are given in eV and momentum matrix elements in eV.$\textrm{\AA}$. Mo states refer to energy bands in the MoSe$_2$ layer, 1W to the first WSe$_2$ layer (adjacent to MoSe$_2$), and 2W to the second WSe$_2$ layer. For conduction band states, the $\pm$ label indicates the energetic ordering of the spin-split bands.}
\begin{centering}
\begin{tabular}{cccccccccccccc}
\hline
\hline
 &  &  & \multicolumn{2}{c}{H$^{\textrm{h}}_{\textrm{h}}$} &  & \multicolumn{2}{c}{H$^{\textrm{h}}_{\textrm{h}}$/H$^{\textrm{M}}_{\textrm{h}}$} &  & \multicolumn{2}{c}{H$^{\textrm{h}}_{\textrm{h}}$/H$^{\textrm{X}}_{\textrm{h}}$} &  & \multicolumn{2}{c}{H$^{\textrm{h}}_{\textrm{h}}$/H$^{\textrm{h}}_{\textrm{h}}$}\tabularnewline
 \hline
$v$ & $c$ &  & Energy & $\left|p_{vc}^{\alpha}\right|,\alpha$  &  & Energy & $\left|p_{vc}^{\alpha}\right|,\alpha$  &  & Energy & $\left|p_{vc}^{\alpha}\right|,\alpha$  &  & Energy & $\left|p_{vc}^{\alpha}\right|,\alpha$ \tabularnewline
\hline
Mo  & Mo$-$ &  & 1.4272 & 4.2392,$+$ &  & 1.4307 & 4.3023,$+$ &  & 1.4292 & 4.2796,$+$ &  & 1.4356 & 4.3692,$+$\tabularnewline
Mo & Mo$+$ &  & 1.4473 & 0.2045,$z$ &  & 1.4507 & 0.1300,$z$ &  & 1.4491 & 0.1989,$z$ &  & 1.4555 & 0.1909,$z$\tabularnewline
1W & 1W$-$ &  & 1.3840 & 0.5735,$z$ &  & 1.3898 & 0.6196,$z$ &  & 1.3735 & 0.6213,$z$ &  & 1.3836 & 0.7641,$z$\tabularnewline
1W & 1W$+$ &  & 1.4441 & 5.7063,$-$ &  & 1.4502 & 5.7071,$-$ &  & 1.4415 & 5.6618,$-$ &  & 1.4520 & 5.6401,$-$\tabularnewline
2W & 2W$-$ &  & --- & --- &  & 1.3848 & 0.4331,$z$ &  & 1.3740 & 0.4779,$z$ &  & 1.3806 & 0.4526,$z$\tabularnewline
2W & 2W$+$ &  & --- & --- &  & 1.4325 & 5.7795,$+$ &  & 1.4285 & 5.2790,$+$ &  & 1.4347 & 5.6930,$+$\tabularnewline
\hline
\hline
1W & Mo$-$ &  & 1.1149 & 0.1695,$-$ &  & 1.1345 & 0.1700,$-$ &  & 1.1380 & 0.1810,$-$ &  & 1.1410 & 0.1708,$-$\tabularnewline
1W & Mo$+$ &  & 1.1349 & 0.6214,$+$ &  & 1.1545 & 0.6356,$+$ &  & 1.1578 & 0.6824,$+$ &  & 1.1609 & 0.6486,$+$\tabularnewline
2W & Mo$-$ &  & --- & --- &  & 1.1602 & 0.0091,$-$ &  & 1.1572 & 0.0536,$z$ &  & 1.1470 & 0.2336,$+$\tabularnewline
2W & Mo$+$ &  & --- & --- &  & 1.1802 & 0.0179,$+$ &  & 1.1771 & 0.0048,$+$ &  & 1.1668 & 0.1058,$z$\tabularnewline
\hline
\hline
\end{tabular}
\end{centering}
\label{tab:Hhh}
\end{table}

\begin{table}[h!]
\caption{Same as in Table~\ref{tab:Hhh} but for the $38.2 \degree$ structures.}
\begin{centering}
\begin{tabular}{ccccccccccc}
\hline
\hline
 &  &  & \multicolumn{2}{c}{$38.2 \degree$} &  & \multicolumn{2}{c}{$38.2 \degree$/R$^{\textrm{X}}_{\textrm{h}}$} &  & \multicolumn{2}{c}{$38.2 \degree$/R$^{\textrm{M}}_{\textrm{h}}$}\tabularnewline
 \hline
$v$ & $c$ &  & Energy & 
$\left|p_{vc}^{\alpha}\right|,\alpha$ &  & Energy & $\left|p_{vc}^{\alpha}\right|,\alpha$ &  & Energy & $\left|p_{vc}^{\alpha}\right|,\alpha$\tabularnewline
\hline
Mo & Mo$-$ &  & 1.4411 & 4.5747,$+$ &  & 1.4429 & 4.5757,$+$ &  & 1.4429 & 4.5748,$+$\tabularnewline
Mo & Mo$+$ &  & 1.4617 & 0.1073,$z$ &  & 1.4635 & 0.1303,$z$ &  & 1.4636 & 0.1323,$z$\tabularnewline
1W & 1W$-$ &  & 1.3802 & 0.3629,$z$ &  & 1.3750 & 0.3859,$z$ &  & 1.3898 & 0.3734,$z$\tabularnewline
1W & 1W$+$ &  & 1.4297 & 5.7411,$+$ &  & 1.4250 & 5.6921,$+$ &  & 1.4407 & 5.7479,$+$\tabularnewline
2W & 2W$-$ &  & --- & --- &  & 1.3914 & 0.4273,$z$ &  & 1.3751 & 0.3479,$z$\tabularnewline
2W & 2W$+$ &  & --- & --- &  & 1.4409 & 5.7710,$+$ &  & 1.4238 & 5.7227,$+$\tabularnewline
\hline
\hline
1W & Mo$-$ &  & 1.1116 & 0.0233,$+$ &  & 1.1101 & 0.0161,$+$ &  & 1.1315 & 0.0177,$+$\tabularnewline
1W & Mo$+$ &  & 1.1322 & 0.0029,$z$ &  & 1.1308 & 0.0023,$z$ &  & 1.1521 & 0.0024,$z$\tabularnewline
2W & Mo$-$ &  &  &  &  & 1.1754 & 0.00060,$z$ &  & 1.0442 & 0.0012,$-$\tabularnewline
2W & Mo$+$ &  &  &  &  & 1.1960 & 0.00003,$-$ &  & 1.0648 & 0.0005,$+$\tabularnewline
\hline
\hline
\end{tabular}
\end{centering}
\end{table}

\begin{figure}[h!]
 \centering
 \includegraphics{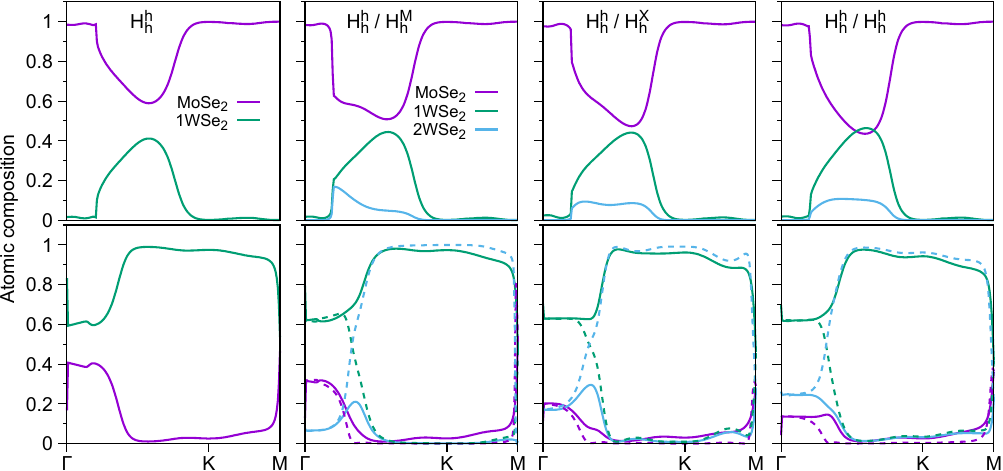}
 \caption{Calculated atomic composition for HBL and HTL systems with H-type stacking ($60 \degree$). Top row: lowest conduction band. Bottom row: top valence bands (the second highest valence band present in HTL systems is shown with dashed lines).}
 \label{fig:Hhh_atomic}
\end{figure}

\begin{figure}[h!]
 \centering
 \includegraphics{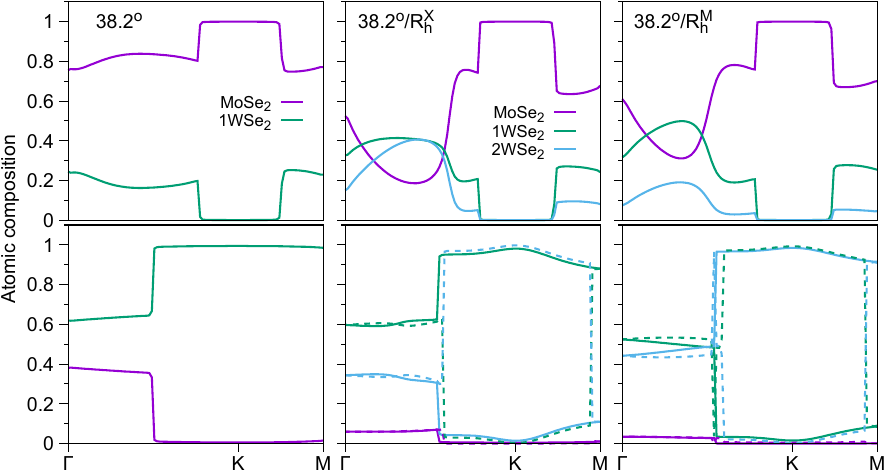}
 \caption{Same as in Fig. S~\ref{fig:Hhh_atomic} but for the structures with $38.2 \degree$ twist angle between MoSe$_2$ and WSe$_2$, and R-type stacking between the two WSe$_2$ layers.}
 \label{fig:38.2_atomic}
\end{figure}


\clearpage

\newpage

\section{Interlayer K-K exciton binding energies}

Interlayer exciton electrostatic potential:

\begin{equation}
V(k)=\frac{e^{2}}{\Lambda\varepsilon_{0}k\varepsilon(k)}
\end{equation}

\begin{figure}[h!]
\centering
\includegraphics{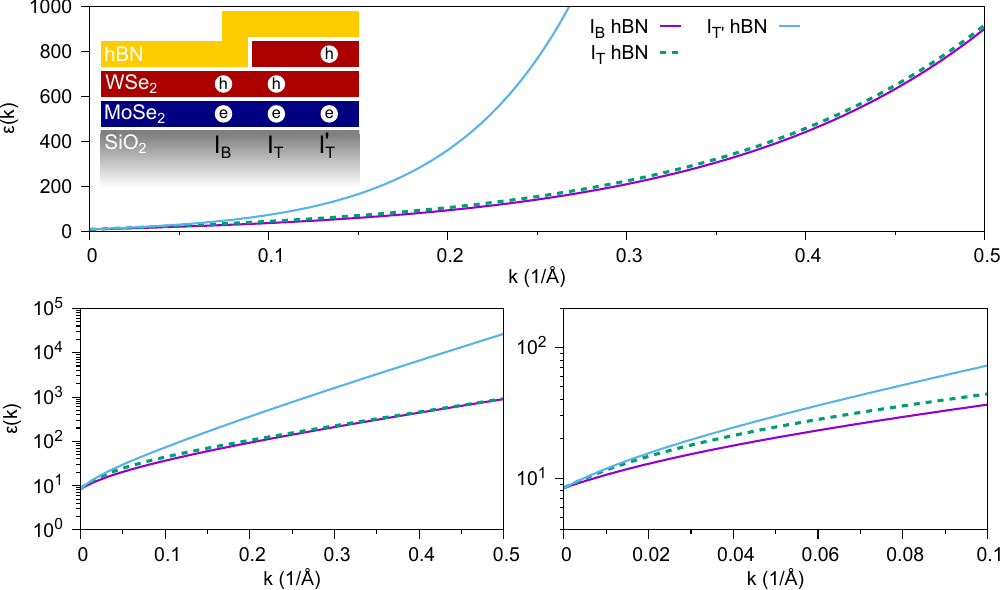}
\caption{Calculated dielectric constant for the different interlayer excitons present in the HBL and HTL structures.}
\label{fig:excitons}
\end{figure}

\begin{table}[h!]
\caption{Effective masses for energy bands and reduced exciton masses ($\mu$). The index $\pm$ indicates the energy splitting of conduction bands in MoSe$_2$.}
\begin{centering}
\begin{tabular}{cccccc}
\hline
\hline
 & vW & cMo$+$ & cMo$-$ & $\mu_{+}$ & $\mu_{-}$\tabularnewline
 \hline
HBL & 0.39 & 0.62 & 0.54 & 0.2394 & 0.2265\tabularnewline
HTL & 0.38 & 0.67 & 0.58 & 0.2425 & 0.2296\tabularnewline
\hline
\hline
\end{tabular}
\end{centering}
\end{table}

\begin{table}[h!]
\caption{Exciton binding energies.}
\begin{centering}
\begin{tabular}{cccccc}
\hline
\hline
$\textrm{I}_{\textrm{B}+}$ & $\textrm{I}_{\textrm{B}-}$ & $\textrm{I}_{\textrm{T}+}$ & $\textrm{I}_{\textrm{T}-}$ & $\textrm{I}_{\textrm{T}+}^{\prime}$ & $\textrm{I}_{\textrm{T}-}^{\prime}$\tabularnewline
\hline
103.35 & 101.42 & 103.80 & 101.90 & 95.15 & 93.54\tabularnewline
\hline
\hline
\end{tabular}
\end{centering}
\end{table}

